\documentclass[preprint,
 amsmath,amssymb,
 aps,
prl
]{revtex4-2}

\usepackage{dsfont}
\usepackage{pstricks}
\usepackage[tight]{subfigure}
\usepackage{verbatim}
\usepackage{units}
\usepackage{multirow}
\usepackage{enumitem}
\usepackage{mathrsfs}
\usepackage{leftidx}
\usepackage{xspace}
\usepackage{braket}
\usepackage{bbm}
\usepackage{overpic}
\usepackage{amsfonts,amssymb,amsmath}
\usepackage{graphicx}
\usepackage{dcolumn}
\usepackage{xcolor}
\usepackage{bm}

\begin{document}

\preprint{revisiting 2/3, 3/5}

\title{Revisiting the Physics of Hole-Conjugate Fractional Quantum Hall Channels}


\author{D. C. Glattli}
\email{christian.glattli@cea.fr}
 
\author{C. Boudet}%
\author{A. De}%
\author{P. Roulleau}%

\affiliation{%
 Nanoelectronics group, Université Paris-Saclay, CEA,
 CNRS, SPEC, 91191 Gif-sur-Yvette cedex, France
}%

\date{\today}

\begin{abstract}
%
We revisit the physics of hole-conjugate Fractional Quantum Hall (FQH) phases characterized by counter-propagating edge channels at filling factors above 1/2. We propose a minimal and intuitive model that successfully accounts for all experimentally observed features, introducing a paradigm shift in the understanding of hole-conjugate edge channel dynamics. Our model identifies inter-channel charge equilibration as the sole essential mechanism, eliminating the need to invoke charge modes or upstream neutral modes, as posited in prior theoretical frameworks. By incorporating fictitious reservoirs along the edge, the model qualitatively and quantitatively reproduces key observations, including counterintuitive upstream effects previously misattributed to neutral modes. We provide predictions for electrical and thermal conductance as well as current noise for filling factors 2/3 and 3/5. Additionally, we address the case of non-dissipative reservoirs, which preserve conductance properties while exhibiting infinite thermal relaxation lengths.
\end{abstract}

\maketitle

\section{\textbf{INTRODUCTION}}
The Fractional Quantum Hall Effect (FQHE) arises in 2D electron system (2DES) under high perpendicular magnetic field $B$, at low temperature, when the filling factor $\nu=en_{s}h/B$ value is fractional \textemdash $\ n_{s}$ being the electron density. At simple rational values of $\nu$, the 2DES forms topologically insulating states.
Conduction occurs on the edge via topological chiral edge channels. For $\nu=p/(2p+1)$, i.e. $1/3$, $2/5$, $3/7$, ..., there are p co-propagating edge channels. For $\nu=p/(2p-1)$, with $p\geq 2$, counter-propagating edge channels occur. Early theoretical work \cite{MacDonald1990, Johnson1991} modelled the $\nu=2/3$ edge by an outer chiral integer edge channel, conductance $e^2/h$ and an inner counter-propagating fractional edge channel with conductance  $(-1/3) \times (e^2/h$). However, this picture gives a total conductance of $4/3 \times (e^2/h$) while all observations find $2/3 \times (e^2/h$). Similarly the edge channels of the hole-conjugate FQH phase at bulk filling factor $\nu=3/5$ is described by a downstream integer channel and two upstream fractional edge channels with conductance $(-1/5) \times (e^2/h)$ leading to a conductance $7/5 \times (e^2/h)$ while all observations find $3/5 \times (e^2/h)$. Solving this puzzle, the Kane-Fisher-Polchinski (KFP) model \cite{Kane1994, Kane1995} showed, in a Chiral Luttinger Liquid (CLL) picture,  that the combination of Coulomb coupled counter-propagating modes with spatially random disorder on the edges leads to a downstream chiral charge mode with conductance $2/3$ and a bosonic upstream neutral mode. The KFP approach was later exploited by the same authors to calculated the thermal conductance $\kappa$ of hole-conjugate FQH edge channels \cite{Kane1997}. They show that the thermal conductance of any elementary fractional channel with conductance $(1/m)e^2/h$, $m$ odd integer, is equal to the thermal conductance  quantum $\kappa_{0}T=(\pi^{2}/3)(e^{2}/h)k_{B}T$, where $k_{B}$ is the Boltzmann constant and $T$ the temperature. As a direct consequence, the thermal conductance of the $\nu=2/3$ is expected to be $0\times \kappa_{0}T$ and that of the $\nu=3/5$ edge is counter-intuitively found negative and equal to $-1\times \kappa_{0}T$. 
Later, experiments led by the Weizmann group using current noise measurements \cite{Bid2010} found evidence of upstream excitation signatures, therefore assigned to the predicted neutral mode although we will provide another interpretation here.
The puzzling physics of the $2/3$ and $3/5$ edges has then led to a flurry of theoretical models \cite{Takei2011, Takei2015, Protopopov2017, Nosiglia2018, Park2019, Spaanslatt2020, Fujisawa2021, Ponomarenko2023, Manna2023} inspired by the KFP model to improve it or to adapt it to situations where more complex reconstruction of the edge occurs \cite{Meir1994, Wang2013}. Concurrently experimental works based on conductance, noise, or thermal conductance have explored the nature of the edge \cite{Gross2012, Altimiras2012,  Inoue2014, Banerjee2017, Rosenblatt2020, Srivastav2021, LeBreton2022, Nakamura2023, Hashisaka2023}. Recently, artificial counter-propagating $2/3$ edge channels have been realised by coupling left and right regions of different ($1$ and $1/3$) filling factor \cite{Cohen2019, Hashisaka2021, Hashisaka2023}. 

However the KFP model, although widely accepted, lacks of physical intuition for non-experts. It is difficult to separate what comes from the bosonic modes entering in the CLL picture with what comes from the disorder which mixes the counter-propagating modes by tunnelling. Also, it becomes clear from a study of the propagation of neutral and charge modes that neutral modes cannot propagate beyond a micron size charge equilibration length, see \cite{Fujisawa2021} and the Appendix D, figure 6. It is then unlikely that the neutral modes, emitted $40\mu m$ away from the Quantum Point Contact detector, can be the source of the upstream excitation signatures observed in \cite{Bid2010}. In the present work, we will present a new intuitive and powerful approach to model hole-conjugate edge channels, which solves this issue and yields predictions compatible with all observations. We show that charge equilibration between counterpropagating channels is the only relevant ingredient. Invoking charge and neutral modes is not necessary.

\begin{figure}[ht]
    \includegraphics[width=15.5cm]{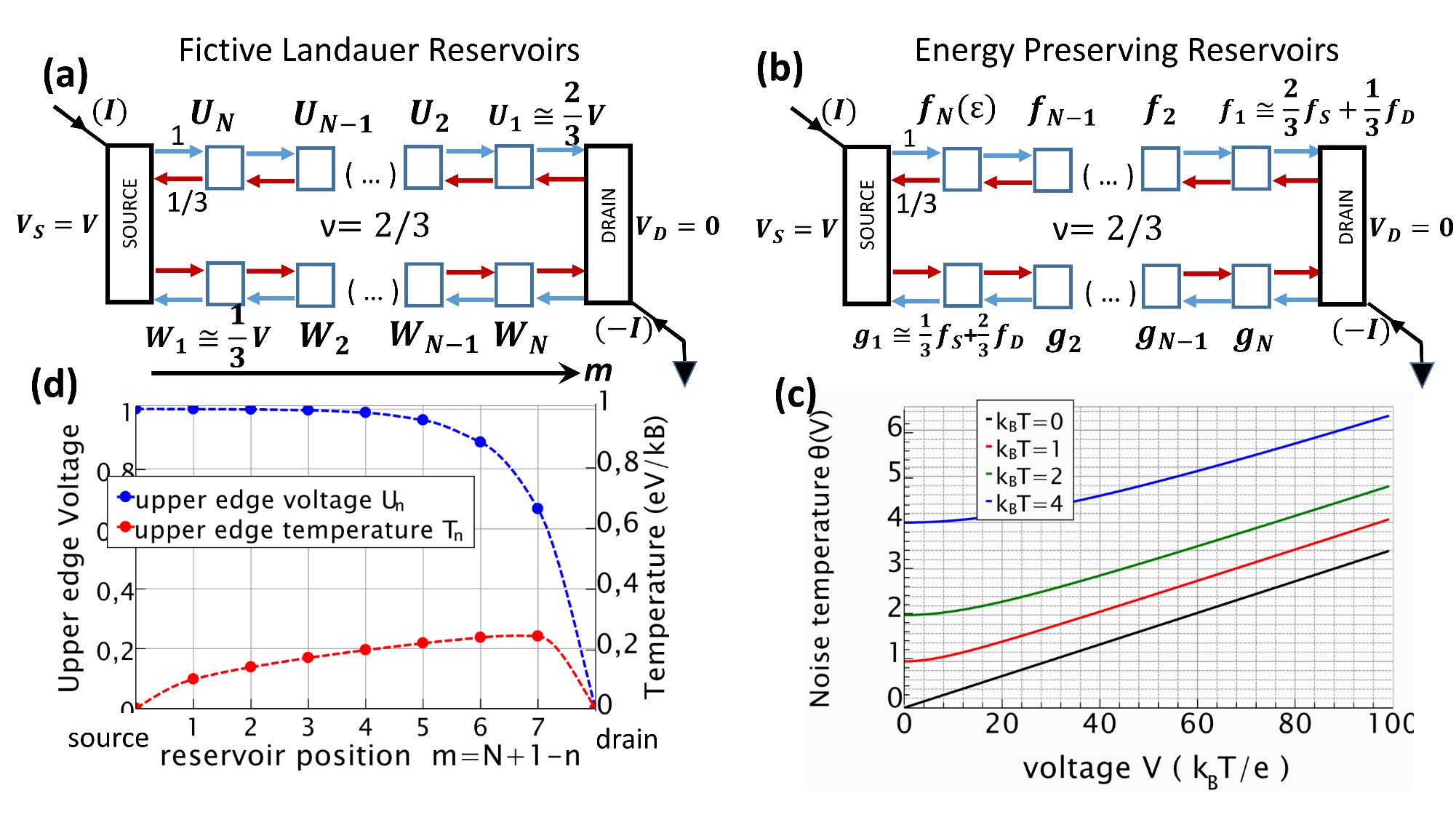}
    \caption{(a) Hall bar with $N$ fictive Landauer reservoirs inserted on upper and lower edge. (b) same but with energy preserving reservoirs. (c) Landauer reservoirs: noise temperature of the Hall bar versus voltage for several  drain-source contacts temperature. The noise is computed for $N=7$ LRs. (d) Landauer reservoirs: spatial voltage (blue) and temperature (red) distribution along the upper edge for $N=7$. $m=N+1-n$ denotes the spatial position from left to right.}
    \label{fig:Fig-1}
   
\end{figure}

\section{\textbf{THE MODEL}}
Here we provide a novel approach using basic tools of mesoscopic physics. The model discussed in the following should be better viewed as a toy model. 
It is not intended to provide a microscopic description of channel mixing; it is intended to explore, in an easy way, what the minimal ingredients are that lead to results compatible with observations. 

We consider a Hall bar, with physical source and drain ohmic contacts on the left and right ends, see Fig.\ref{fig:Fig-1}(a) and (b). 
To enable charge equilibration between counter-propagating channels, N fictitious charge reservoirs are inserted along the upper and lower edges. The use of ohmic contacts  has been recently considered in \cite{Staebler2022,Staebler2023} to study the heat flow along quantum Hall edge, however we do not consider capacitive here charging effects. We concentrate, as a first step, only on the charge equilibration properties of the reservoirs. We consider two generic cases schematically shown in Fig.\ref{fig:Fig-1}. In Fig.\ref{fig:Fig-1}(a), the reservoirs are fictive ohmic contacts described as Landauer reservoirs (LR). In Fig.\ref{fig:Fig-1}(b) the reservoirs are Energy-Preserving Reservoirs (EPR), as introduced in \cite{deJong1996}. In both cases, the $1$ and $-1/3$ outer and inner channels ballistically propagate between two consecutive reservoirs. We recall that for an LR, quasiparticles are absorbed whatever their energy and are emitted following a Fermi distribution having the temperature and the chemical potential of the reservoir (here expressed in voltage units). LRs thus provide full equilibration of both charge and energy between the counterpropagating channels. An opposite situation occurs for EPRs: an excitation hitting the reservoir at energy $\epsilon$ is absorbed and re-emitted at the same energy, and charge equilibration alone occurs. Note that, using reservoirs, we will not address a third possible regime: coherent channel mixing \cite{Acciai2022, Hashisaka2021}. 

In the following, we find that both LR and EPR lead to the same conductance values for $N$ reservoirs, rapidly converging with $N$ towards the KFP $2/3 \times (e^2/h)$ conductance fixed point. For LR we also recover the zero and negative thermal conductance respectively predicted for $\nu=2/3$ or $3/5$ by the KFP model and observed in \cite{Altimiras2012, Banerjee2017, LeBreton2022}. We emphasize that the case of energy-preserving channel equilibration was overlooked in the literature, and our result shows that dissipative processes are not necessary to reach the KFP conductance fixed point. However, beyond conductance, comparing LRs and EPRs leads to quantitatively different temperature distribution and current noise. 

\section{\textbf{CONDUCTANCE AND CURRENT NOISE}}

We first concentrate on predictions for conductance and current noise for a $2/3$ Hall bar for both LRs and EPRs. 
 
 For LRs, the evenly distributed fictive contacts are floating at potential $U_{n}$ ($(W_{n}$) along the upper (lower) edge with no external current entering them. For convenience we set the conductance quantum $G_{0}=e^{2}/h$ equal to unity. The Landauer-Büttiker relation For the $n^{th}$ fictive reservoir on the upper edge channel is:
 \vspace*{-.1in}
 \begin{equation}
     \label{Landauer}
     0=\frac{4}{3}U_{n}-U_{n+1}-\frac{1}{3}U_{n-1}
     \vspace*{-.1in}
 \end{equation}
 A similar relation holds for the potentials $W_{n}$ of the lower edge. For energy preserving reservoirs:
  \vspace*{-.1in}
  \begin{equation}
     \label{Beenakker}
     0=\frac{4}{3}f_{n}(\epsilon)-f_{n+1}(\epsilon)-\frac{1}{3}f_{n-1}(\epsilon)
      \vspace*{-.1in}
 \end{equation} with $f_{n}(\epsilon)$ the non-equilibrium energy distribution of the $n^{th}$ EPR of the upper edge channel. A similar set of equations holds for the energy distributions $g_{n}(\epsilon)$ of the lower edge channel.
 Here we must clarify the use of energy distribution $f_{n}(\epsilon)$ for the -1/3 fractional channel. As one considers only perfect ballistic transmission between reservoirs, the amount of current carried in the energy interval $d\epsilon$ is $-e^{*}f_{n}(\epsilon)d\epsilon$, $e^{*}=e/3$, is consistent with the $-e^{2}/3h$ inner channel conductance. We think, treating the distribution as Fermi-like and disregarding chiral Luttinger Liquid aspects is, at this stage, a reasonable simplification not qualitatively changing the results below.  
 
 Using appropriate Landauer-Büttiker boundary conditions on source and drain contacts at potential $V_{S}$ and $V_{D}$ respectively, we find the following voltage distribution for LRs (see Appendix A):
 \vspace*{-.1in}
 \begin{equation}
     \label{LR-voltage}
     U_{n}=\frac{V_{S}(1-e^{-n\lambda })+V_{D}(e^{-n\lambda}-e^{-(N+1)\lambda})}{1-e^{-\lambda (N+1)}}
      \vspace*{-.1in}
 \end{equation} with $\lambda=\ln{3}$.
 Similarly, the energy distribution for the EPRs is:
  \vspace*{-.1in}
  \begin{equation}
     \label{EPR-distribution}
     f_{n}(\epsilon)=\frac{f_{S}(\epsilon)(1-e^{-n\lambda })+f_{D}(\epsilon)(e^{-n\lambda }-e^{-(N+1)\lambda})}{1-e^{-\lambda (N+1)}}
 \end{equation}
with $f_{S(D)}(\epsilon)$ the Fermi distributions of the source (drain) contacts at temperature $T$ and voltage $V_{S}$($V_{D})$ respectively. For the lower edge channel, exchanging source and drain voltages (energy distributions) in equations \ref{LR-voltage}(\ref{EPR-distribution}) leads to similar expressions for $W_{n}$(or $g_{n}(\epsilon)$). The conductance for $N$ reservoirs is found to be:
 \vspace*{-.1in}
\begin{equation}
    \label{conductance}
    G_{N}=\frac{2}{3}\frac{1+\frac{1}{3}e^{-\lambda N}}{1-\frac{1}{3}e^{-\lambda N}}
     \vspace*{-.1in}
\end{equation}
 for both LRs and EPRs. In general, as long as conductance properties are concerned, both dissipative and non-dissipative reservoirs remarkably give the same conductance. The first five values of $G_{N}$ when $N$ goes from $0$ to $N=4$ are respectively: 
$2(2/3)$, $\frac{5}{4}(2/3)$, $\frac{14}{13}(2/3)$, $\frac{41}{40}(2/3)$, and $\frac{122}{121}(2/3)$. With just two fictive reservoirs the $2/3$ KFP conductance fixed point is reached within $7.7\% $ accuracy. 

It is interesting to compare the current flowing through the upper and lower edge channel. For the upper edge we find a \textit{downstream} current converging towards $(2/3)e^{2}V/h$:

\begin{equation}
    I_{downstr}=\frac{2}{3}V\frac{1}{1-e^{-(N+1)\lambda}}
     \vspace*{-.1in}
\end{equation} 
while for the lower edge we find an \textit{upstream} current exponentially vanishing with $N$:

\begin{equation}
    I_{upstr}=\frac{2}{3}V\frac{e^{-(N+1)\lambda}}{1-e^{-(N+1)\lambda}}
     \vspace*{-.1in}
\end{equation}

\vspace{1in}
We can repeat the same approach for the 3/5 FQHE edge which is made of an outer (1) downstream integer channel an two inner (-1/5) upstream fractional channels. For LR, Eq.\ref{Landauer} now becomes:
\begin{equation}
     \label{Landauer3-5}
     0=\frac{7}{5}U_{n}-U_{n+1}-\frac{2}{5}U_{n-1}
 \end{equation}
 and similarly in the case of EPR, Eq.\ref{Beenakker} becomes: 
 \begin{equation}
     \label{Beenakker3-5}
     0=\frac{7}{5}f_{n}(\epsilon)-f_{n+1}(\epsilon)-\frac{2}{5}f_{n-1}(\epsilon)
 \end{equation}

 For the computation of LR voltages and EPR energy distribution, equations \ref{LR-voltage} and \ref{EPR-distribution} hold with now $\lambda = 5/2$. The conductance for $N$ reservoirs and 3/5 Hall bar becomes:
\begin{equation}
    \label{conductance3-5}
    G_{N}=\frac{3}{5}\frac{1+\frac{2}{5}e^{-\lambda N}}{1-\frac{2}{5}e^{-\lambda N}}
\end{equation}

We now concentrates on the case of dissipative Landauer reservoirs to investigate dissipation and noise at $\nu=2/3$. For simplicity we set the voltages $V_{S}=V$ and $V_{D}=0$. Along the upper channel, from left to right, we observe from Eq.\ref{LR-voltage} that the voltages $V_{n}$ remain almost equal to $V_{S}=V$ all along the channel  and drops rapidly to $V_{D}=0$ near the drain contact, see Fig.\ref{fig:Fig-1}(d). Reciprocally, from right to left, the lower channel voltage remains almost zero and rapidly rise to $V$ near the source. This feature has several implications: we expect strong dissipation near the drain (source) for the upper (lower) edge and almost no dissipation along the Hall bar.  Indeed we find that the power dissipated in the n$^{th}$ upper edge fictive contact is (see Appendix A): $P_{n}=\frac{1}{2}(\frac{4}{3}V)^{2}e^{-2\lambda n}/(1-e^{-(N+1)\lambda})^{2}$. For the lower edge, the power $Q_{n}=P_{n}$. This confirms the existence of "hot spots" near the drain and the source for respectively the upper and lower edge channel, as found in previous theoretical models \cite{Takei2011,Nosiglia2018, Park2019} . To check that the calculated power expressions make sense, we have computed the total  power $P$ in the Hall bar, summing  the drain, the source, and the upper and lower edge contributions. We find a total power $P=I.V$ as it should be. Knowing the set of $P_{n}$, we can calculate the temperature distribution $T_{n}$ ($T'_{n}$) for the upper (lower) edge channel using  the universal quantum thermal conductance $\kappa_{0}$. Equal thermal conductance for elementary $1$ and $-1/3$ chiral channels was predicted in \cite{Kane1997}. Fixing the temperature of the physical drain and source reservoirs equal to $T$, the temperature $T_{n}$ of the fictive reservoirs is obtained by solving the standard heat flow equations (see Appendix A). The solution is:
 \vspace*{-.10in}
\begin{equation}
    \label{temperatureTn}
   (k_{B} T_{n})^{2}=(k_{B}T)^2+\frac{3}{(2\pi)^{2}}(eV)^{2}[ 1-n/(N+1)-e^{-2\lambda n}]
\end{equation}
The same temperature distribution $T'_{n}=T_{n}$ is found for the lower edge. For the upper edge , from right to left, we observe a square root like increases of the spatial temperature distribution followed by a sudden drop to $T$ near the drain, see Fig.\ref{fig:Fig-1}(d). A similar symmetric spatial variation occurs for the lower edge. Importantly, the temperature increase along the edge channel, signals a diffusive \textit{upstream} heat flow contrasting with the downstream charge flow. As our model does not require introducing a specific neutral collective mode, the counter-propagating heat flow plays its role. Our model thus provides a new perspective on the nature of the neutral mode. We will show later that the upstream heat flow explains the upstream signatures observed in experiment \cite{Bid2010}. We emphasizes that for $\nu=2/3 $ the upsream heat flow occurs because of the finite voltage bias inducing dissipation in each fictive reservoir. In contrast, in absence of current, section IV will show that the thermal conduction is diffusive and non-chiral as found in \cite{Kane1997}.

Knowing the temperature distribution along the edge of the $2/3$ Hall bar, we now can compute the non equilibrium hot electron thermal noise for $V\ne 0$. To proceed, we consider all independent thermal current noise sources generated between each pair of fictive reservoirs. For the pair $(n,n-1)$ on the upper edge the current noise spectral density is $S_{I}^{n,n-1}=2(k_{B}T_{n}+\frac{1}{3}k_{B}T_{n-1})$ with the $T_{n}$ given by Eq.\ref{temperatureTn}. A fluctuation $\delta I_{n,n+1}$ between a $(n,n+1)$ fictive contact pair  is found giving rise to the fluctuation $\delta I_{S}^{n,n+1}=\frac{2}{9}\delta I_{n,n+1}e^{-(N-1-n)\lambda}/(1-e^{-(N+1)\lambda})$  at the source contact and, from current conservation, an equal fluctuation in the drain contact. Adding the uncorrelated mean square fluctuations of the noise of all pairs $(n,n+1)$ provides the total drain current noise. Setting source and drain temperatures to $T=0$ for clarity, we find the total current noise power, can be expressed  as:
 \vspace*{-.1in}
\begin{equation}
    S_{I_{S}}=F_{LR}2eI
\end{equation}
with the (fake) Fano factor $F_{LR}$ given by :
\begin{equation}
\label{FanoLR}
    F_{LR}=\frac{\sqrt{3}}{\pi\sqrt{N+1}}\{\frac{1}{3}+\frac{2}{27}\sum_{m=2}^{N}e^{-2(m-1)\lambda}(\sqrt{m}+\frac{1}{3}\sqrt{m-1}) \}
\end{equation}
Summation (\ref{FanoLR}) is valid for large $N$ and gives $F_{LR}\simeq 0.19298203.../ \sqrt{(N+1)}$. For large lengths, the non-equilibrium hot electron thermal noise vanishes like the square root of $\sqrt{(N+1)}$, the Hall bar length. Similar length dependence was predicted using different theoretical approaches \cite{Park2019, Park2019-b} and experimentally observed \cite{Hashisaka2023}. This is in agreement with the diffusive heat flow leading to zero thermal conductance for infinite length predicted in \cite{Kane1997} for $\nu=2/3$. We observe from the summation in Eq.\ref{FanoLR}  where $m$ stands for $m=N-n$, that the main noise contributions do not  comes from the "hot spots" (near the drain for the upper edge) but from noise sources situated close to the source having the smallest noise temperature $T_{n} $  ( $n$ close to $N$). 
For finite temperature $T$, the asymptotic linear variation of the noise with applied voltage $V$ becomes accurate for $V > \simeq 35 k_{B}T $ see Fig.\ref{fig:Fig-1}(c). 

We now turn to the dissipation and the current noise for the case of energy preserving reservoirs instead of LRs. Because of the nature of the fictive EPRs, we do not expect any dissipation along the upper and lower channels except at source and drain which are physical LRs. Indeed, the sum of the power dissipated in the source and the drain is found equal to the Joule power $IV$, see Appendix B. Thus, contrasting with the Landauer reservoir case where the thermal and charge equilibration lengths were found equal, the thermal length is here infinite. These two opposite thermal length behaviours have been experimentally observed \cite{Rosenblatt2020, Kumar2022, Melcer2022}. For EPR, we expect a current noise different from the LR case. 
The non-equilibrium energy distribution $f_{n}$ and $g_{n}$  given by Eq.\ref{EPR-distribution}, are occupation probabilities taking values between zero and one. This leads to  fluctuations of the state occupation number at energy $\epsilon$ equal to $f_{n}(\epsilon)(1-f_{n}(\epsilon))$.  This results in the current noise spectral density associated to a $(n+1;,n)$ EPR pair equal to  $S_{I}^{{n+1,n}}=2\frac{e^2}{h}[k_{B}\Theta_{n+1} +\frac{1}{3}k_{B}\Theta_{n}]$: 
where we have defined the fictive temperatures $k_{B}\Theta_{n} =\int_{-\infty}^{\infty} d\epsilon f_{n}(\epsilon)(1-f_{n}(\epsilon)) $
\begin{table}[ht]
    \centering
    \begin{tabular}{|c|c|c|c|c|} \hline 
         $N$&  $F_{LR}$& $F_{LR}\sqrt{N+1}$&$F_{EPR}$&$ 3^{N}.F_{EPR}$ \\ \hline 
    
         5&  0,07911& 0,1937& 0,002698&0,6556 \\\hline 
         6&  0,07305& 0,1933& 0,000908& 0,6621\\ \hline 
         7&  0,06827& 0,1931& 0,000304&0,6649\\ \hline
         $\infty$&  0& 0.19298...& 0&2/3\\ \hline
    \end{tabular}
    \caption{Fake Fano factors for different lengths $N$. For small $N=5$-$7$, $F_{LR}$ is not obtained from the large $N$ expression Eq.(\ref{FanoLR}) but from exact calculation at finite $N$. }
     \vspace*{-.15in}
    \label{tab:Fano}
\end{table}

The total noise observed in the drain or source can be calculated as we did above for LRs, replacing $T_{ n}$ by $\Theta_{n}$. In contrast with the LR case, the noise is found exponentially vanishing with the size $N$ (see Appendix B). The fictive temperatures growing linearly with $V$, we can define the fake Fano factor $F_{EPR}$. For large $N$ we found $F_{EPR}\simeq 0.666 (1/3)^{N}$ contrasting with
$F_{LR}\propto N^{-1/2}$. Such exponential dependence was reported in \cite{Melcer2022}. Table \ref{tab:Fano} gives Fano factor values for some finite $N$ values.
\begin{figure}[ht]
    \vspace*{-.1in}
    \centering
    \includegraphics[width=12.6cm]{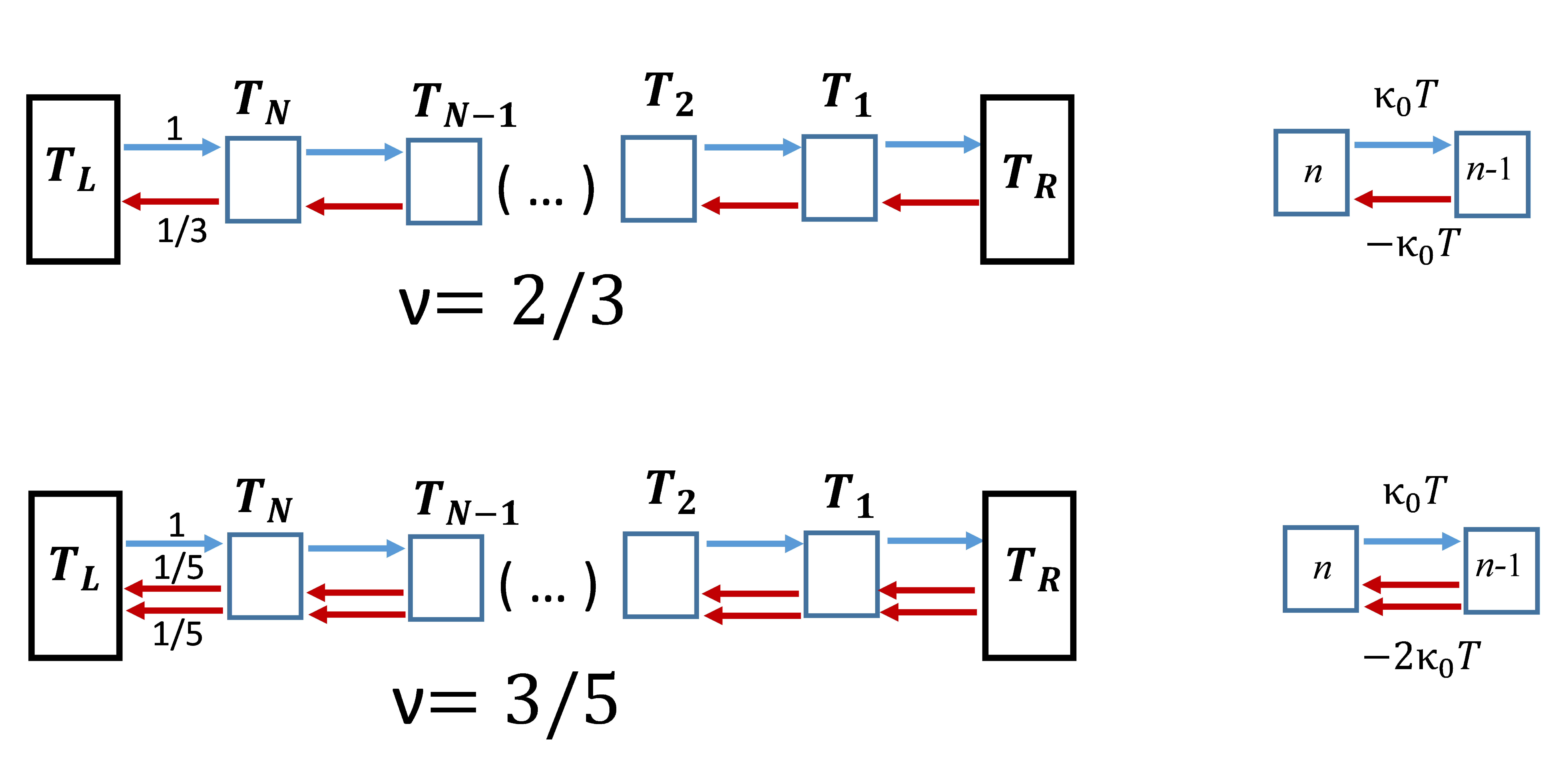}
    \vspace*{-.1in}
    \caption{ thermal conductance. On the top is schematically represented the 2/3 upper edge with the chiral (1) and -1/3) conductance channels on the left and the $\kappa_{0}T$ and $-\kappa_{0}T$ thermal conductance channels on the right. The total thermal conductance is 1-1=0. On the bottom figures a similar schematic drawing of the conductance and thermal edge channels are represented for $\nu=3/5$. The total thermal conductance is 1-1-1=-1 corresponding to upstream heat flow}
     \vspace*{-.1in}
    \label{fig:Fig-2-Heat}
\end{figure}

\section{\textbf{THERMAL CONDUCTANCE}}

We now address the thermal conductance at $\nu=2/3$ and $\nu=3/5$. As the prediction of thermal conductances in these two regimes, confirmed by observations, was an important check of the KFP model, it is interesting to see if our model reproduces these predictions. Only the case of LR will be considered here as, EPR exchanging energy but no heat,  thermal conductance is not defined.  To proceed, we consider a situation were no voltage bias is applied between the left and right ohmic contacts. Instead different temperatures are applied between left and right contacts with temperature $T_{L}$ an $T_{R}$ respectively. The quantity of interest here is no longer the current but the heat flow. 

We consider the thermal flow through the upper edge. Similar calculation hold for the lower edge. 
Contrary to the above calculation, as there is no current, no dissipation occurs in any fictive reservoir. The temperature $T_{n}$ of the $n^{th}$ reservoir is expected between $T_{L}$ and $T_{R}$. Expressing that the sum of the heat entering and exiting the $n^{th}$ reservoir cancels, we get for $\nu=2/3$: 
\begin{equation}
    \label{2-3-heatbalance}
    0=2T_{n}^2-T_{n-1}^2-T_{n+1}^2
\end{equation}
The solution is of the form $T_{n}^{2}=A+Bn$ with $A$ and $B$ determined by the boundary conditions $T_{n=0}=T_{R}$ and $T_{n=N+1}=T_{L}$. This gives
\begin{equation}
    \label{2-3-temperature}
    T_{n}=\sqrt{T_{R}^2+(T_{L}^2-T_{R}^2)n/(N+1)}
\end{equation}
The heat flow is $\Dot{Q}=\kappa_{0}(T_{L}^{2}-T_{R}^{2})/2(N+1)$. For $\nu=2/3$ the thermal conduction is thus found diffusive. Chiral features are absent and the thermal conductance vanishes in the large $N$ limit as $1/\sqrt{N+1}$. This agrees with the predictions of reference \cite{Kane1997} based on the KFP approach. 

In contrast for $\nu=3/5$, reference \cite{Kane1997} predicts a surprising ballistic-like thermal conductance given by $-\kappa_{0}T$, i.e. corresponding to upstream heat flow. In the present model, which is based on a series of Landauer reservoirs randomising charge and energy between them, it seems unlikely that a ballistic-like thermal conductance can appear. However, this is what we are going to show below. Writing the heat balance equation for $\nu=3/5$ yields:
\begin{equation}
    \label{3-5-heatbalance}
    0=3T_{n}^2-T_{n+1}^2-2T_{n-1}^2
\end{equation}
 The complete solution is $T_{n}^{2}=A+B2^{n}$, with $A$ and $B$ being determined by the boundary conditions for $n=0$ and $n=N+1$. This gives the distribution of the square temperature along the edge:
\begin{equation}
    \label{3-5-temperature}
    T_{n}^{2}=\frac{T_{R}^{2}-2^{-N-1}T_{L}^{2} + 2^{n-N-1}(T_{L}^{2}-T_{R}^{2})}{1-2^{-N-1}}
\end{equation}

The heat flow going from reservoir $n$ to reservoir $n-1$ is $\Dot{Q}_{n}=(\kappa_{0}/2)(T_{n}^{2}-2T_{n-1}^2)$. Its expression is:
\begin{equation}
    \label{3-5-heatflow}
    \Dot{Q}_{n}=-(\kappa_{0}/2)\frac{T_{R}^2-T_{L}^{2} 2^{-(N+1)}}{1-2^{-(N+1)}}
\end{equation}
The heat flow does not depends on $n$ as expected for energy conservation and is found  to flow upstream. For large $N$ we get $-\kappa_{0}T_{R}^{2}/2$. This is equivalent to the ballistic heat flow of single ballistic upstream channel in perfect agreement with reference \cite{Kane1997}.

\section{\textbf{QUANTUM POINT CONTACT CONDUCTANCE AT $\nu=2/3$}}

\begin{figure}[ht]
    \vspace*{-.1in}
    \centering
    \includegraphics[width=12.6cm]{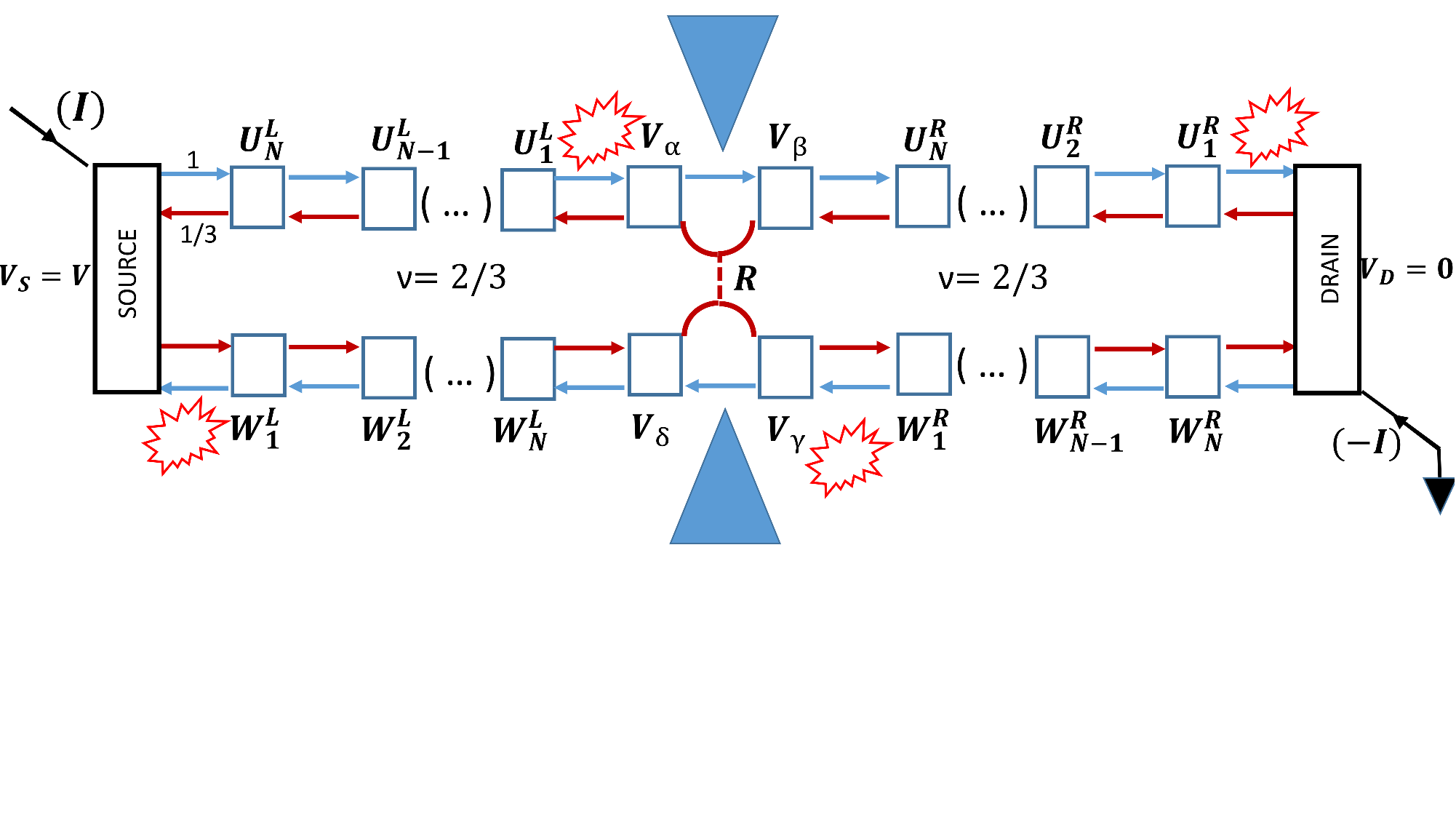}
    \vspace*{-.1in}
    \caption{ 2/3 Hall bar with a QPC in its middle. $N$ LRs are inserted along the edges on each side of the QPC. In red are indicated the "hot spots" where most dissipation occurs.}
     \vspace*{-.1in}
    \label{fig:Fig-2}
\end{figure}

We now address  practical situations
exploiting the present model. While we have considered above an homogeneous Hall bar, we now investigate the conductance when a quantum point contact is located in the middle of the Hall bar and induces tunnelling between the upstream upper and lower fractional $(-1/3)$ inner edge channels while the downstream integer channel is transmitted. To proceed, 
we introduce $N$ fictive reservoirs on the left side of the QPC (both for the lower and upper edge) and, symmetrically, $N$ reservoirs on the right side, see Fig.\ref{fig:Fig-2}. We seek for a solution of the  upper and lower  potentials $U_{n}^{L(R)}$ and $W_{n}^{L(R)}$ respectively on the right (R) and left (L) side of the QPC. The boundary conditions, given by Landauer-Büttiker relations at the source and the drain and at the QPC, see Appendix C, provide a determination of the unknown potentials. 
For large $N$, the QPC conductance is found:
\vspace*{-.1in}
\begin{equation}
    G=\frac{2}{3}\frac{e^2}{h}\frac{1+\mathcal{R}/2}{1+\mathcal{R}}
    \label{QPCconductance}
     \vspace*{-.1in}
\end{equation}
where $\mathcal{R}$ is the reflection probability of the fractional inner channel. Note that, as we consider partitioning a fractional channel, Luttinger liquid corrections may apply and we should better consider $\mathcal{R}$ as the ratio of the backscattered current to the current incoming on the QPC.
 The graph of QPC conductance versus $\mathcal{R}$ is given in Appendix C. Interestingly, for full reflection we get the half conductance quantum $\frac{e^2}{2h}$. A similar result was found in \cite{Nakamura2023} in a model attempting to explain the observed $0.5$ conductance plateau and in the theoretical works of reference \cite{Manna2023, Manna2024}. We find the same conductance $G(\mathcal{R})$ for energy-preserving reservoirs instead of LRs, confirming that conductance alone can not distinguish between dissipative and non-dissipative regime. 
 
 It is interesting to look at the potential distribution along the edges for the LR case. Starting from the source contact at $V$, the upper edge voltage remains almost equal to $V$ but quickly drops near QPC to reach $V_{\alpha}=V\frac{1+\mathcal{R}/2}{1+\mathcal{R}}$. Then on the upper right of the QPC, starting at $V_{\beta}=V_{\alpha}$ the voltages remain almost constant and rapidly drop near the drain at $V_{D}=0$. Similarly, starting from the drain, the lower edge voltages remain almost zero and rapidly increase near the QPC to reach $V_{\gamma}=V\frac{\mathcal{R}/2}{1+\mathcal{R}}$. On the left of the QPC, the voltages remain almost equal to $V_{\delta}=V_{\gamma}$ to finally rapidly increase near the source to $V$. The rapid drops of potentials being associated with dissipation we can identify two new "hot spots" on both side of the QPC as shown in Fig.\ref{fig:Fig-2}.

 \section{\textbf{DELTA-T NOISE VERSUS NEUTRAL MODE PARTITION NOISE}} 
\begin{figure}[ht]
    \centering
    \includegraphics[width=14.5cm]{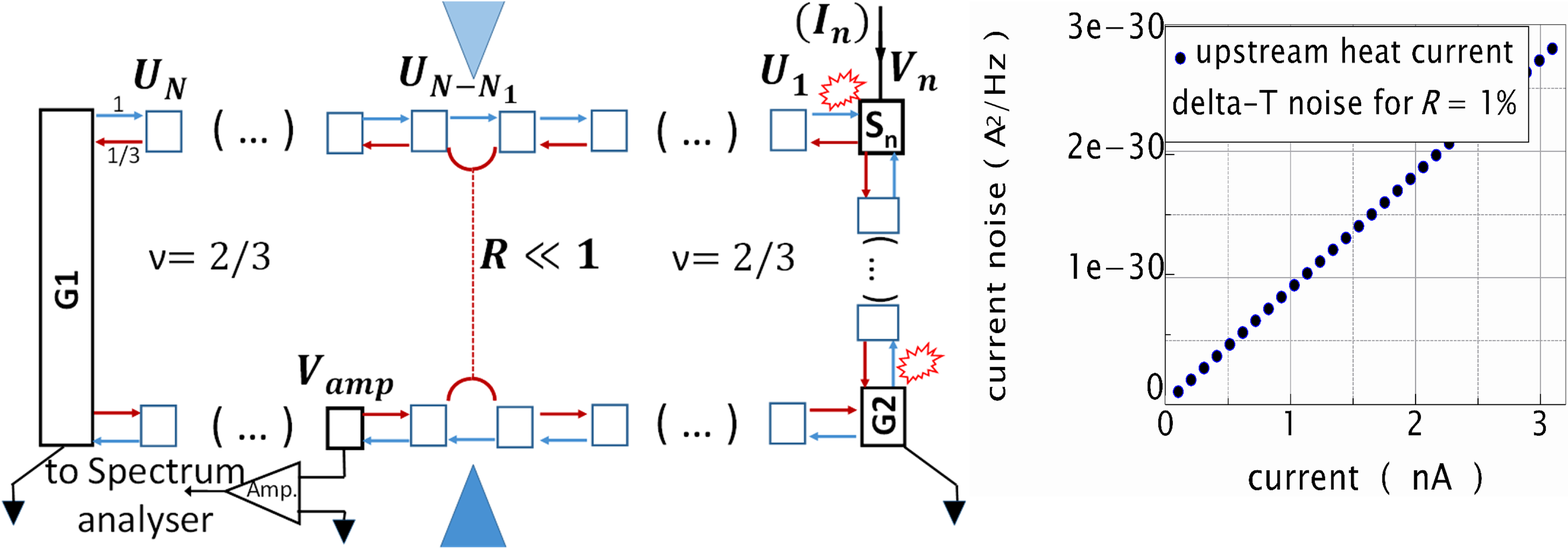}
    \caption{ Left: set-up similar to Bid's experiment \cite{Bid2010}. Weak reflection $\mathcal{R}<<1$ is considered. The upper unpstream heat flow heats the upper side of the QPC. Right: calculated delta-T noise.}
    \label{fig:Fig-3}
\end{figure}

 As another application of the model,  we now consider the set-up of Fig.\ref{fig:Fig-3}, which is similar to Bid's experiment reporting counter-propagating neutral modes \cite{Bid2010}. This seminal experiment had a strong impact on the community as it was the first to report the manifestation of upstream effects. However, exploiting the results of \cite{Fujisawa2021}, it is clear that neutral modes are unable to propagate more than very few charge relaxation lengths \textit{at all frequencies}. A calculation of the neutral and charge mode attenuation length is shown in Fig.\ref{fig:Fig-S3}. Then, it is unlikely that the neutral mode can be the source of partition noise in experiment \cite{Bid2010} where the distance between the source of neutral modes is about 40 times the charge equilibration length leading to an attenuation factor of $4.10^{-18}$. A reinterpretation of the experimental observation is therefore necessary. Here we show that the thermal gradient due to the counter-propagating heat flow found above, section III, is responsible for the observed noise. 
 In the set-up of Fig.\ref{fig:Fig-3}, the source contact ($S_{n}$), located at the upper right of a QPC, is at potential $V_{n}$ and injects the current $I_{n}$. The left and right contacts ($G_{1}$) and ($G_{2}$) of the Hall bar are grounded. A fourth physical contact ($V_{amp}$) on the lower left side of the QPC records the voltage fluctuations $S_{V_{amp}}$, from which the current noise $S_{I}=S_{V_{amp}}(2e^{2}/3h)^{2}$ can be inferred. 

 For simplicity, $k_{B}T=0$ and the QPC weakly backscatters the (-1/3) inner edge ($\mathcal{R}<<1$) leaving the Hall bar almost unperturbed. The Hall bar edges are modelled by $N>>1$ fictive LRs. According to the above results, the current $I_{n}$ flows from ($S_{n}$) to contact ($G_{2}$). A hot spot forms at ($G_{2}$) and on the left side of ($S_{n}$). This generates an \textit{upstream heat flow} flowing along the upper edge and heating the upper side of the QPC. For $N_{1}$ LRs between the QPC and ($G_{1}$), the temperature $T_{up}$ of the upper channel at the QPC is, according to Eq. \ref{temperatureTn}:
  \vspace*{-.1in}
 \begin{equation}
     k_{B}T_{up}=\frac{\sqrt{3}}{2\pi} eV_{n} \sqrt{\frac{N_{1}}{N}}
     \label{Tup}
      \vspace*{-.079in}
 \end{equation}

In contrast, the lower channel, connected to zero temperature zero voltage contacts remains at $T=0$. This situation generates a "delta-T" shot noise \cite{Lumbroso2018, Reulet2020, Rech2020, Sassetti2022,Acciai2024} at the QPC whose one side is at $T_{up}$ and the other side at $T=0$. A similar conclusion, but in a full CLL approach was found in \cite{Takei2011}. $T_{up}$ is given by, see \cite{Reulet2020}:
 \vspace*{-.1in}
\begin{equation}
    S_{I}=2(1/3)(\mathcal{R}\frac{2e^2}{3h})\ln{(2)}k_{B}T_{up}
     \vspace*{-.079in}
\end{equation}
where the (1/3) factor come from the tunnelling of quasiparticle of charge $e^{*}=e/3$. For $N_{1}=N/2$, zero temperature and transmission $0.99$ ($\mathcal{R}=1\%$), $I_{n}=3$nA gives $S_{I}\simeq 2.8 10^{-30}$$A^{2}$/Hz a value close to the reported experimental value. If LRs are now replaced by EPRs, the distribution function $f_{N-N_{1}}$ at the location of the QPC is close to the $T=0$ Fermi distribution of contact $G_{1}$. Therefore \textit{no delta-T noise} is expected. 

\section{SUMMARY}
To conclude, the present model provides a novel, intuitive, and powerful way to obtain the conductance, the noise, the thermal conductance and more general counter-intuitive features of hole-conjugate FQH edge channels. We have addressed particularly the $2/3$ and $3/5$ fractional edge. The model also predicts the anomalous $0.5e^{2}/h$ QPC conductance plateau at $\nu=2/3$. The model suggest a paradigm shift in our understanding of the edge physics: taking into account inter-channel charge equilibration is the only requirement to get all the striking and counter-intuitive features of hole-conjugate FQHE. Charge and neutral modes can be fully ignored as a first step.  When the conductor is biased by a finite current, the model gives a new interpretation of neutral excitations as a counter-propagating heat flow detectable by its delta-T noise. A Bid type experiment \cite{Bid2010} observing noise would signal a dissipative equilibrium regime (i.e. LRs) generating Delta-T noise. In contrast, observing no noise would signal no thermal equilibration (i.e. EPRs).  Finally, to get a better quantitative description, charge and neutral collective modes can be reintroduced  as a second step  on top of the model in order to take into account the high frequency dynamics of the edge \cite{Fujisawa2021, Lin2021, Hashisaka2023} and chiral Luttinger effects. 

We thank Ankur Das, Yuval Gefen, Moshe Goldstein, Sourav Manna, Carles Altimiras, Olivier Maillet, François Parmentier for useful discussions. We acknowledge the European Union H2020 research and innovation program under grant agreement No. 862683, “UltraFastNano”.

\vspace{3cm}

\vspace{0.75cm}
\textbf{APPENDIX A} 

\textbf{Hall bar conductance for $N$ fictive LRs:}

To compute the Hall bar conductance with LRs inserted between source and drain, we use Landauer-Büttiker relations. The boundary conditions are: a current $I$ sent by the external circuit enters the physical source contact and accordingly, a current $-I$ enters the drain contact. The Voltages of the source (drain) contacts are $V_{S}$ ($V_{D}$) respectively. The fictive LR voltages are noted $U_{n}$ ($W_{n}$) for upper (lower) edges. We have:
\begin{subequations}
\label{eq:whole}
\begin{eqnarray}
I=\frac{4}{3}V_{S}-\frac{1}{3}U_{N}-W_{1}
\label{subeq:2}
\end{eqnarray}
\begin{equation}
0=\frac{4}{3}U_{n}-U_{n+1}-\frac{1}{3}U_{n-1}
\label{subeq:4}
\end{equation}
\begin{equation}
0=\frac{4}{3}W_{n}-W_{n+1}-\frac{1}{3}W_{n-1}
\label{subeq:3}
\end{equation}
\begin{equation}
-I=\frac{4}{3}V_{D} -U_{1}-\frac{1}{3}W_{N}
\label{subeq:1}
\end{equation}
\end{subequations}
with $n$ running from $1$ to $N$ following a counterclockwise (upstream) order. Seeking for a solution $\propto \exp(-n\lambda)$ we find $\lambda=0$ or $\ln(3)$. The general solution is a linear combination of these two solutions whose coefficient are determined by the source and drain boundary conditions \ref{subeq:2},\ref{subeq:1}, which leads to: 
\begin{equation}
     \label{LR-voltage-S}
     U_{n}=\frac{V_{S}(1-e^{-n\lambda })+V_{D}(e^{-n\lambda}-e^{-(N+1)\lambda})}{1-e^{-\lambda (N+1)}}
 \end{equation} with $\ln(\lambda)=3$. Using Eq.\ref{subeq:1} (or Eq. \ref{subeq:2}) we find the conductance $G_{N}=(2/3)\frac{1+(1/3)^{(N+1)}}{1-(1/3)^{(N+1)}}$ for N reservoirs, including the case $N=0$. A similar computation can be done for $\nu=3/5$, with $\lambda=5/2$, and will not be repeated here.

 To compute the temperature of each floating fictive reservoirs, we set the voltages $V_{S}=V$ and $V_{D}=0$ for simplicity and we first calculate the Joule power $P_{n}$ dissipated in each reservoir:    
     \begin{align}
     P_{n} & =\frac{1}{2}[(U_{n+1}-U_{n})^{2}+\frac{1}{3}(U_{n}-U_{n-1})^{2}]  \nonumber \\
     & = \frac{1}{2}(\frac{4}{3}V)^2 e^{-2n\lambda}/(1-e^{-(N+1)\lambda})^{2}
     \label{power-n}
     \end{align}
A similar calculation hold for the $n^{th}$ reservoir of the lower edge with power $Q_{n}=P_{n}$. To check that the power calculated makes sense, we compute the total power$P$ dissipated in the Hall bar in the limit $N\to\infty$. 
The Joule power dissipated at the drain and the source by the upper and the lower edge channels are respectively : $P_{D}=P_{S}=\frac{1}{2}(V(1-e^{-\lambda})^{2}$  
\begin{equation}
    \label{totalpower}
    P=P_{D}+P_{S}+2\sum_{n=1}^{N} P_{n}=V^{2}((\frac{2}{3})^{2}+\frac{1}{8}(\frac{4}{3})^{2})=IV
\end{equation}
as it should be.

To obtain the temperature distribution along the edge, we have to solve the following heat flow equation:
\begin{equation}
    \label{solvingtemperature}
    P_{n}=\frac{\kappa_{0}}{2}(T_{n}^{2}-T_{n-1}^{2}+T_{n}^{2}-T_{n+1}^{2})
\end{equation}
As $P_{n}\propto e^{-2n\lambda}$, the general solution can be written as $T_{n}^2=Ae^{-2n\lambda} +Bn + C$ where $A$, $B$, and $C$ are determined from the drain and source thermal boundary conditions. The solution, valid for large $N$, is given by Eq. \ref{temperatureTn} in the main text. For small $N$, values of $T_{n}$ are given in table  \ref{tab:temperature} for $k_{B}T=0$ and $V=1$. 

\begin{table}[ht]
    \centering
    \begin{tabular}{|c|c|c|c|c|c|c|c|} \hline 
         N&  $T_{1}$&  $T_{2}$&  $T_{3}$&  $T_{4}$& $T_{5}$ &  $T_{6}$&$T_{7}$ \\ \hline 
         4&0,2297  &0,2121  & 0,1747 & 0,1237 &  &  & \\ \hline 
         5& 0,2345 & 0,2232 & 0,1949 & 0,1593 & 0,1126 &  & \\ \hline 
         6&0,2382  &0,2310  &0,2082 &0,1805  &0,1474  &0,1042  & \\ \hline 
         7&0,2409  & 0,2367 & 0,2177 &0,1949  &0,1688  & 0,1378 &0,0974 \\ \hline 
        
    \end{tabular}
    \caption{Distribution of temperature $T_{n}$ for the $n^{th}$ Landauer reservoir $n$, for $N=4-7$. The $T_{n}$ are in calculated for unit voltage $V$ and  $T=0$}
    \label{tab:temperature}
\end{table}
Once we have the set of $T_{n}$ values, we can compute the (hot) thermal (chiral) current noise source $S_{I}^{n+1,n}=2k_{B}[T_{n+1}+\frac{1}{3}T_{n}]$ between pair of reservoirs (n+1,n) on the upper edge and calculate, for each independent pairs, the induced source and drain fluctuations. A similar calculation is done for the lower edge. We found that the current noise $S_{I_{S}}^{(n+1,n)}$ viewed at the source contact due to the (n+1,n) pair is : $S_{I_{S}}^{(n+1,n)}=S_{I}^{n+1,n}\left( \frac{2}{9}\frac{e^{-(N-n-1)\lambda}}{(1-e-{(N+1)\lambda }}\right) ^{2}$. Adding all independent current noise of each pairs for both upper and lower edge and adding the drain and source contacts contributions, we get the total noise of the Hall bar. It is interesting to note the following. The hottest temperature for the upper (lower) edge occurs close to the drain (source), forming  "hot spots".\ref{fig:Fig-1} but the current noise contribution is exponentially suppressed by a factor$\simeq e^{-2N\lambda}$. On the contrary the current noise sources close to the source (drain) contact contribute to with the largest weight while they have the lowest noise temperature. This corresponds to "noise spots". However, here no shot noise occurs in the present Toy model, only thermal noise with temperature $T_{n}$ rising linearly with $V$. As the temperatures $T_{n}$ near the source scale as $1/\sqrt{N+1}$ the non-equilibrium noise vanishes for large length. 

We thus get the effective noise temperature $\tilde{\Theta}_{N}(V)$ of the whole 2/3 Hall bar defined as: $S_{I}^{N}=4G_{N}k_{B} \tilde{\Theta}_{N} (V)$, for N fictive probes when a drain-source voltage $V$ s applied. For zero drain and source temperature, $\tilde{\Theta}_{N}(V) \propto V$. To characterise the noise, we define the effective (or fake) Fano factor by $2eI=2eG_{N}V$, i.e. $F_{LR}=2k_{B}\tilde{\Theta} (V)/eV$.

Table \ref{tab:Fano}, in the main text, gives values of the fake Fano factor. The values for N=5-7 result from an exact numerical calculation for $N$ reservoirs. We see that for $N\leq 5$ the large $N\to\infty$ limit given by Eq.\ref{FanoLR} is almost reached.

\vspace{0.75cm}
\textbf{APPENDIX B}

\textbf{Hall bar conductance for $N$ fictive EPRs:}

The calculation of the unknown distributions $f_{n}$ and $g_{n}$ for the $n^{th}$ reservoir of the upper and lower edge is very similar to the calculation of potentials for LRs. The boundary conditions are the Fermi distribution $f_{S}(\epsilon)$ and $f_{D}(\epsilon)$ at respectively the source and drain reservoirs. We get the set of following linear equations for the upper edge:
\begin{subequations}
\label{eq:distributionfunction}
\begin{eqnarray}
0=\frac{4}{3}f_{N}(\epsilon)-f_{S}(\epsilon)-\frac{1}{3}f_{N-1}(\epsilon),
\label{subeq:distr1}
\end{eqnarray}
\begin{equation}
0=\frac{4}{3}f_{n}(\epsilon)-f_{n+1}(\epsilon)-\frac{1}{3}f_{n-1}(\epsilon), \text{\space \space  $1\leq n \leq N-2$,}
\label{subeq:distr2}
\end{equation}
\begin{equation}
0=\frac{4}{3}f_{1}(\epsilon)-f_{2}(\epsilon)-\frac{1}{3}f_{D}(\epsilon).
\label{subeq:distr3}
\end{equation}
\end{subequations}
A similar set of equation holds for lower edge.
The general solution is:
\begin{equation}
f_{n}(\epsilon)=(1-\frac{e^{\lambda (N+1-n)}-1}{e^{\lambda (N+1)}-1})f_{S}(\epsilon)+\frac{e^{\lambda (N+1-n)}-1}{e^{\lambda (N+1)}-1}f_{D}(\epsilon)
\label{eq: fn}
\end{equation}
for $1\leq n \leq N$, where, as above $\lambda=ln(3)$.

The non-equilibrium distribution $f_{n}$ are thus linear combination of the source and drain Fermi distributions. In the limit of large $N$, $f_{1} \rightarrow \frac{2}{3}f_{S}+\frac{1}{3}f_{D}$, $f_{N}\rightarrow f_{S}$. Similarly, 
$g_{1} \rightarrow \frac{1}{3}f_{S}+\frac{2}{3}f_{D}$ and $g_{N}\rightarrow f_{D}$.
The current at the source is :
\begin{equation}
\begin{split}
I_{D}& =\int_{-\infty}^{\infty}\left(\frac{4}{3}f_{S}(\epsilon)-\frac{1}{3}f_{N}(\epsilon)-g_{1}(\epsilon)  \right)d\epsilon \\
& = \frac{2}{3}\int_{-\infty}^{\infty}\left(f_{S}(\epsilon)-f_{D}(\epsilon)  \right)d\epsilon=\frac{2}{3}V
\end{split}
\label{eq:EPRsouceCurrent}
\end{equation}
The conductance is thus insensitive to the nature of the reservoirs, LRs or EPRs.

\begin{figure}[ht]
    \centering
    \includegraphics[width=14.2cm]{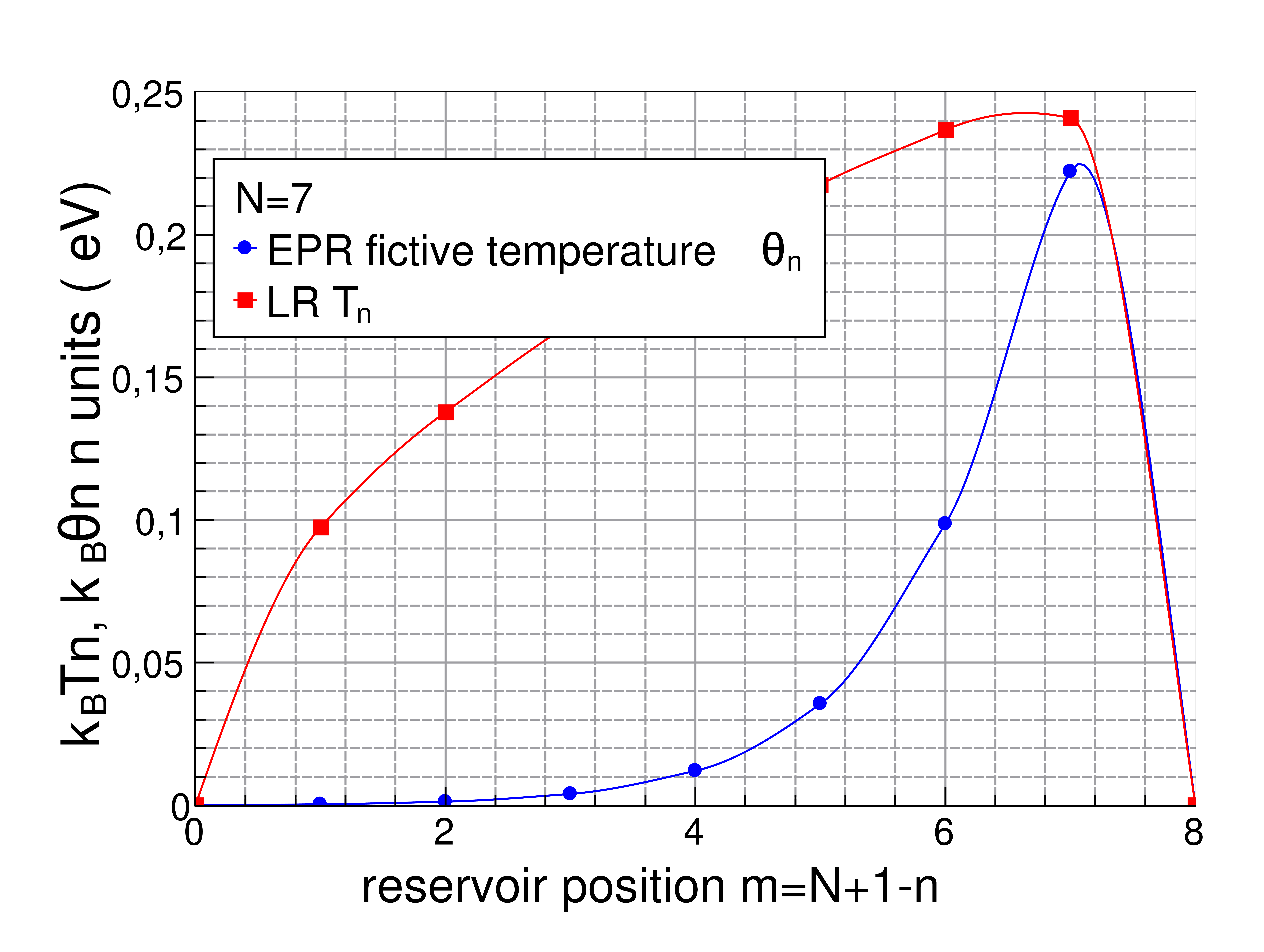}
    \caption{ effective EPR temperatures (blue dots) and non-equilibrium LR temperatures (red dots) calculated for $N=7$. The lines interpolating the points are guides for the eyes}
    \label{fig:Fig-S1}
\end{figure}

We now investigate where dissipation occurs in the 2/3 Hall bar with EPRs. Because of the nature of the fictive BRs we do not expect any dissipation
along the upper and lower channels except at source and the drain which are physical LRs. Indeed, lets calculate the power $P_{S}$ and $P_{D}$ dissipated in the source and in the drain. For simplicity we consider the reservoirs at zero temperature and chose the zero of energy at the chemical potential of the grounded drain contact. For large $N$, we have:
\begin{equation}
\label{energyPreservingPD}
\begin{split}
P_{S}& =\int_{-\infty}^{\infty} d\epsilon (\epsilon-V)[\frac{1}{3}(f_{N}(\epsilon)-f_{S}(\epsilon))+(g_{1}(\epsilon)-f_{D}(\epsilon))] \\
& =\frac{1}{3}\int_{-\infty}^{\infty} d\epsilon (\epsilon-V)(f_{S}(\epsilon)-f_{D}(\epsilon))\\
& =\frac{1}{2}\frac{2}{3}V^{2}
\end{split}
\end{equation}
A similar expression gives $P_{D}=P_{S}$. The total Joule heating is $P=P_{D}+P_{S}=IV$ and thus exclusively occurs in the physical ohmic drain and source contacts. 
The study of dissipation indicates that no generation of thermal current fluctuations can occurs in the fictive EPRs.  However, the distribution $f_{n}$, and $g_{n}$ can generate non thermal population fluctuations. We can define a fictive temperature $\Theta_{n}$ associated wth the distribution $f_{n}$ and defined as:
\begin{equation}
\label{fictive-Tempe}
\begin{split}
\Theta_{n} & =\int_{0}^{eV} d\epsilon f_{n}(\epsilon)(1-f_{n}(\epsilon))\\
& =eV \frac{(1-e^{-n\lambda})(e^{-n\lambda}-e^{-(N+1)\lambda})}{(1-e^{-(N+1)\lambda})^2}
\end{split}
\end{equation}
There is an important difference between the $\Theta_{n}$ for EPRs and the $T_{n}$ for LRs. While the latter have to obey the heat flow equation \ref{solvingtemperature}, the former are entirely defined by the $f_{n}$.
Figure \ref{fig:Fig-S1} compares EPR and LR temperatures

The noise for the case of EPR reservoirs is computed in a way similar to the LR case, except the $T_{n}$ are now replaced by the $\Theta_{n}$. 
Because of the exponential factor relating the noise of a contact pair $(n+1,n)$ to the current noise it produces at the source $\delta I_{S}^{n,n+1}=\frac{2}{9}\delta I_{n,n+1}e^{-(N-1-n)\lambda}/(1-e^{-(N+1)\lambda})$ reduces further the small EPR noise, the final noise of the Hall bar is found exponentially vanishing with the sample size. Some values are indicated in table  \ref{tab:Fano}  in the main text.

\vspace{0.75cm}
\textbf{APPENDIX C}

\textbf{QPC conductance :}

While the above studies were considering an homogeneous Hall bar,we consider a QPC inducing tunneling between the upper and lower (-1/3) inner edge channels.
We first consider the case of fictive Landauer reservoirs. The computation of conductance for  case of EPRs is very similar and leads to same conductance values. Providing the number of reservoirs on the left side and on the right side are infinite, the QPC location has not to be located right in the middle of the Hall bar. Departure from the conductance calculated below will be observable only if the QPC is very close to the drain or to the source.  
To proceed, we introduce $N$ fictive reservoirs on the left side of the QPC (both for the lower and upper edge) and symmetrically $N$ reservoirs on the right side. We seek for a solution of the  upper and lower  potentials $U_{L(R)}$ and $W_{L(R)}$ respectively on the right (R) and left (L) side of the QPC.  
We present the solution in the limit of very large number of reservoirs. We make use of the following property. Consider a part of a 2/3 edge channel with potential $V_{A}$ on the upstream side and $V_{B}$ on the downstream side and $N$ LRs whose index $n$ rises following the upstream direction. the voltages $V_{n}$ of the $n^{th}$ reservoir is given by 
\begin{equation}
\label{Eq.-generic}
     V_{n}=\frac{V_{A}(1-e^{-n\lambda })+V_{B}(e^{-n\lambda}-e^{-(N+1)\lambda})}{1-e^{-\lambda (N+1)}}.
\end{equation}
     We apply this relation to solve the boundary condition at the drain, the source and at the four contacts ($\alpha, \beta, \gamma, \delta$) surrounding the QPC (see figure 2). Namely, the Landauer-Büttiker relation at the drain is:
\begin{equation}
I=\frac{4}{3}V_{S}-\frac{1}{3}U_{N}^{L}-W_{1}^{L}
\end{equation}
and becomes, using the large $N$ limit of Eq.\ref{Eq.-generic}:
\begin{equation}
I\simeq\frac{4}{3}V_{S}-\frac{1}{3}V_{S}-\left(\frac{2}{3}V_{\alpha}+\frac{1}{3}V_{S}  \right)    
\end{equation}
similarly the Landauer-Büttiker relations applied to contact ($\alpha$) on the upper left of the QPC, gives
\begin{equation}
    \begin{split}
    0& =\frac{4}{3}V_{\alpha}-U_{1}^{L}-\frac{1}{3}\left((1-\mathcal{R})V_{\beta}+\mathcal{R}V_{\delta}    \right)\\
    & =\frac{4}{3}V_{\alpha}-(\frac{2}{3}V_{S}+\frac{1}{3}V_{\alpha})-\frac{1}{3}\left((1-\mathcal{R})V_{\beta}+\mathcal{R}V_{\delta}    \right)
    \end{split}
\end{equation}
where we have use the reflection coefficient $\mathcal{R}$ of the fractional inner and  channel and the large $N$ limit of Eq.\ref{Eq.-generic}
\begin{figure}[ht]
    \centering
    \includegraphics[width=12.2cm]{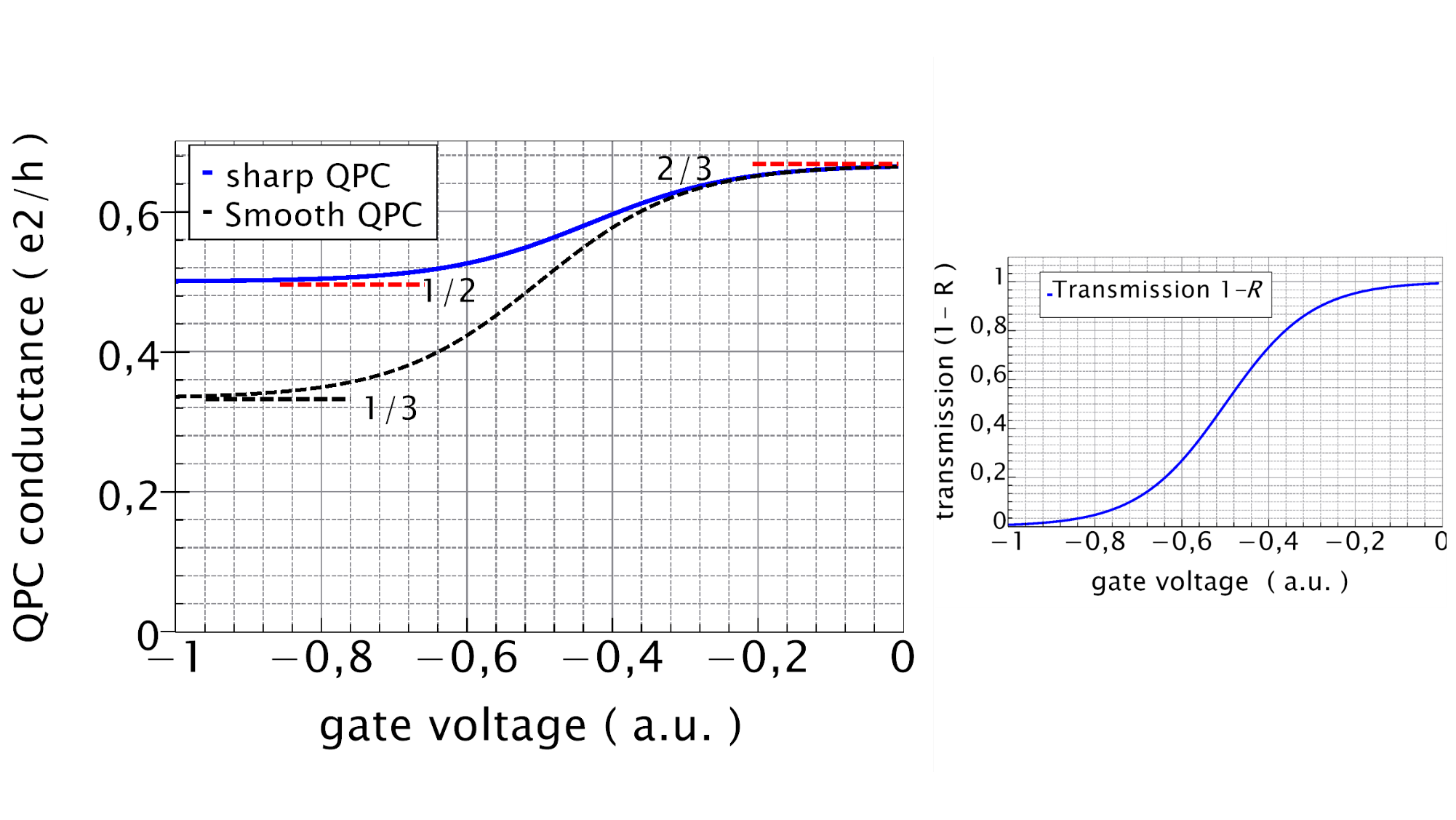}
    \caption{ right figure: QPC conductance versus gate voltage. For sharp edge $G=(2/3)(1+\mathcal{R}/2)/(1+\mathcal{R})$ (blue curve) and for smooth reconstructed edge $G=(1/3)(1+(1-\mathcal{R})$ (black dotted curve). We chose a standard variation of $\mathcal{R}$ versus gate voltage as show in the right figure}
    \label{fig:Fig-S2}
\end{figure}
Repeating the same procedure at contacts $\beta, \gamma$, and $\delta$ and at the drain provides an efficient way to compute the conductance $G=\frac{2}{3}(1+\mathcal{R}/2)/(1+\mathcal{R})$ in presence of the QPC. Figure \ref{fig:Fig-S2} shows the conductance variation. For comparison we have included the case of a smooth edge and smooth QPC for which the 2/3 edge separates into two reconstructed 1/3 edge channels and a $e^{2}/3h$ plateau occurs.
 For full reflection, $\mathcal{R}=1$, the upper edge voltages at the QPC are $V_{\alpha}=V_{\beta}=3V_{S}/4$ while the lower edge voltages are: $V_{\gamma}=V_{\delta}=V_{S}/4$. The voltage drop at the QPC is $V_{QPC}=V_{S}/(1+\mathcal{R})$. This drop could be observable using photo-assisted shot noise which presents a zero temperature noise singularity at the Josephson relation $e^{*}V_{QPC}=hf$, with $f$ the microwave irradiation frequency \cite{XGWen1991, Kapfer2019}. 
 We can reproduce the same findings for energy-preserving reservoirs instead of LRs, introducing the non-equilibrium energy distribution $f_{\alpha}, f_{\beta}, f_{\gamma}$ and $f_{\delta}$ for EPRs around the QPC. As done for voltages in the case of LRs, we use a the generic relation:
 \begin{equation}
\label{Eq.-generic-f}
     f_{n}=\frac{f_{A}(1-e^{-n\lambda })+f_{B}(e^{-n\lambda}-e^{-(N+1)\lambda})}{1-e^{-\lambda (N+1)}}.
\end{equation}
where $f_{A}$ denotes some know distribution on the upstream end of a chain of $N$ EPRs and $f_{B}$ the distribution at the downstream end. Applying this relation to simplify the equations expressing the current conservation at each energy for drain, source and QPC EPRs allows to compute the conductance, which is found identical to the case of LRs.

\vspace{0.75cm}

\textbf{APPENDIX D}

\textbf{neutral mode decay :}
\\
\begin{figure}[ht]
    \centering
    \includegraphics[width=12.2cm]{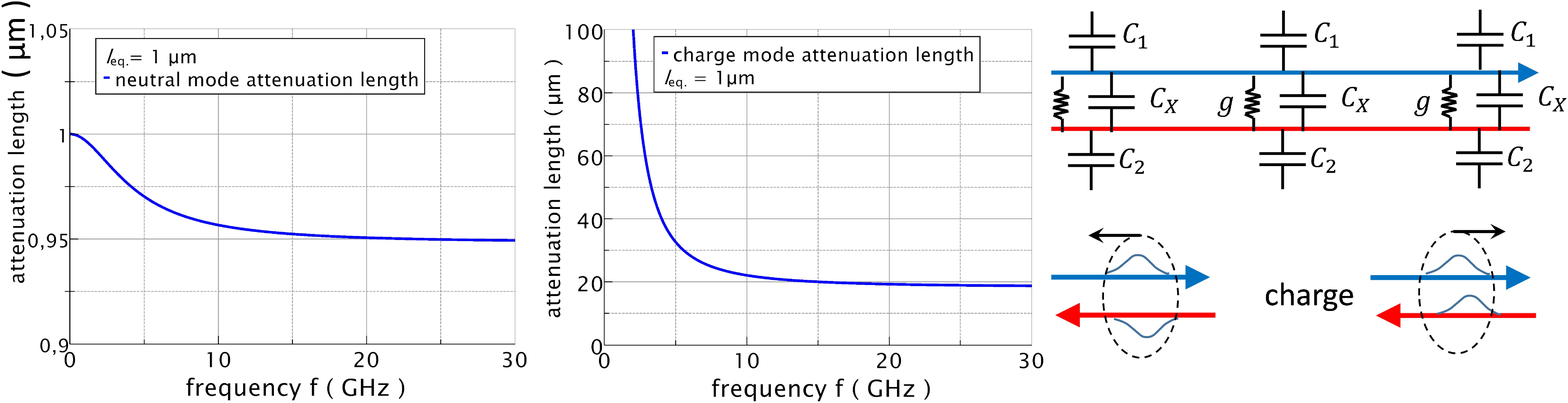}
    \caption{neutral and charge mode attenuation. right figure: capacitive transmission line model, as in \cite{Fujisawa2021}, where distributed capacitances stand for self- and inter-channel interaction. A tunnelling conductance allows for charge equilibration between channels over a length $l_{eq.}$. left figure: neutral mode attenuation length calculated for $l_{eq.}=1\mu$m. Center figure: same for the charge mode.}
    \label{fig:Fig-S3}
\end{figure}
Here we use the dissipative plasmonic transmission line model of counter-propagating 2/3 edge channels as introduced in Ref.\cite{Fujisawa2021}.
Figure \ref{fig:Fig-S3} shows the distributed capacitive model, modelling the interaction with $C_{X}$, $C_{1}$, and $C_{2}$ the capacitance per unit length for the inter-channel coupling and the self-interaction within the outer and inner edge respectively. A uniform inter-channel tunnelling conductance $g=e^{2}/2hl_{eq.}$ is used to mimic the KFP disorder channel mixing, $l_{eq.}$ is the charge mixing length, typically or $\mu$m size. The model is equivalent to the bosonic KFP model and shows upstream neutral and down stream charge modes. The attenuations lengths for neutral and charge modes are shown in Figure \ref{fig:Fig-S3}. We observe that the attenuation length of the neutral mode is smaller than $l_{eq.}$ at all frequencies, while the length is order of magnitude larger for the charge mode and is infinite in the DC limit. The charge mode attenuation has been recently measured in CITE  It is thus unlikely that neutral modes an reach the QPC and generate noise in experiment \cite{Bid2010}, while our suggestion of counter-propagating heat flow yielding a finite temperature difference at the QPC may likely generate the noise observed in this experiment.

\nocite{*}
\bibliography{main-2-3}

\begin{thebibliography}{45}%
\makeatletter
\providecommand \@ifxundefined [1]{%
 \@ifx{#1\undefined}
}%
\providecommand \@ifnum [1]{%
 \ifnum #1\expandafter \@firstoftwo
 \else \expandafter \@secondoftwo
 \fi
}%
\providecommand \@ifx [1]{%
 \ifx #1\expandafter \@firstoftwo
 \else \expandafter \@secondoftwo
 \fi
}%
\providecommand \natexlab [1]{#1}%
\providecommand \enquote  [1]{``#1''}%
\providecommand \bibnamefont  [1]{#1}%
\providecommand \bibfnamefont [1]{#1}%
\providecommand \citenamefont [1]{#1}%
\providecommand \href@noop [0]{\@secondoftwo}%
\providecommand \href [0]{\begingroup \@sanitize@url \@href}%
\providecommand \@href[1]{\@@startlink{#1}\@@href}%
\providecommand \@@href[1]{\endgroup#1\@@endlink}%
\providecommand \@sanitize@url [0]{\catcode `\\12\catcode `\$12\catcode `\&12\catcode `\#12\catcode `\^12\catcode `\_12\catcode `\%12\relax}%
\providecommand \@@startlink[1]{}%
\providecommand \@@endlink[0]{}%
\providecommand \url  [0]{\begingroup\@sanitize@url \@url }%
\providecommand \@url [1]{\endgroup\@href {#1}{\urlprefix }}%
\providecommand \urlprefix  [0]{URL }%
\providecommand \Eprint [0]{\href }%
\providecommand \doibase [0]{http://dx.doi.org/}%
\providecommand \selectlanguage [0]{\@gobble}%
\providecommand \bibinfo  [0]{\@secondoftwo}%
\providecommand \bibfield  [0]{\@secondoftwo}%
\providecommand \translation [1]{[#1]}%
\providecommand \BibitemOpen [0]{}%
\providecommand \bibitemStop [0]{}%
\providecommand \bibitemNoStop [0]{.\EOS\space}%
\providecommand \EOS [0]{\spacefactor3000\relax}%
\providecommand \BibitemShut  [1]{\csname bibitem#1\endcsname}%
\let\auto@bib@innerbib\@empty
\bibitem [{\citenamefont {MacDonald}(1990)}]{MacDonald1990}%
  \BibitemOpen
  \bibfield  {author} {\bibinfo {author} {\bibfnamefont {A.~H.}\ \bibnamefont {MacDonald}},\ }\href {\doibase 10.1103/PhysRevLett.64.220} {\bibfield  {journal} {\bibinfo  {journal} {Phys. Rev. Lett.}\ }\textbf {\bibinfo {volume} {64}},\ \bibinfo {pages} {220} (\bibinfo {year} {1990})}\BibitemShut {NoStop}%
\bibitem [{\citenamefont {Johnson}\ and\ \citenamefont {MacDonald}(1991)}]{Johnson1991}%
  \BibitemOpen
  \bibfield  {author} {\bibinfo {author} {\bibfnamefont {M.~D.}\ \bibnamefont {Johnson}}\ and\ \bibinfo {author} {\bibfnamefont {A.~H.}\ \bibnamefont {MacDonald}},\ }\href {\doibase 10.1103/PhysRevLett.67.2060} {\bibfield  {journal} {\bibinfo  {journal} {Phys. Rev. Lett.}\ }\textbf {\bibinfo {volume} {67}},\ \bibinfo {pages} {2060} (\bibinfo {year} {1991})}\BibitemShut {NoStop}%
\bibitem [{\citenamefont {Kane}\ \emph {et~al.}(1994)\citenamefont {Kane}, \citenamefont {Fisher},\ and\ \citenamefont {Polchinski}}]{Kane1994}%
  \BibitemOpen
  \bibfield  {author} {\bibinfo {author} {\bibfnamefont {C.~L.}\ \bibnamefont {Kane}}, \bibinfo {author} {\bibfnamefont {M.~P.~A.}\ \bibnamefont {Fisher}}, \ and\ \bibinfo {author} {\bibfnamefont {J.}~\bibnamefont {Polchinski}},\ }\href {\doibase 10.1103/PhysRevLett.72.4129} {\bibfield  {journal} {\bibinfo  {journal} {Phys. Rev. Lett.}\ }\textbf {\bibinfo {volume} {72}},\ \bibinfo {pages} {4129} (\bibinfo {year} {1994})}\BibitemShut {NoStop}%
\bibitem [{\citenamefont {Kane}\ and\ \citenamefont {Fisher}(1995)}]{Kane1995}%
  \BibitemOpen
  \bibfield  {author} {\bibinfo {author} {\bibfnamefont {C.~L.}\ \bibnamefont {Kane}}\ and\ \bibinfo {author} {\bibfnamefont {M.~P.~A.}\ \bibnamefont {Fisher}},\ }\href {\doibase 10.1103/PhysRevB.51.13449} {\bibfield  {journal} {\bibinfo  {journal} {Phys. Rev. B}\ }\textbf {\bibinfo {volume} {51}},\ \bibinfo {pages} {13449} (\bibinfo {year} {1995})}\BibitemShut {NoStop}%
\bibitem [{\citenamefont {Kane}\ and\ \citenamefont {Fisher}(1997)}]{Kane1997}%
  \BibitemOpen
  \bibfield  {author} {\bibinfo {author} {\bibfnamefont {C.~L.}\ \bibnamefont {Kane}}\ and\ \bibinfo {author} {\bibfnamefont {M.~P.~A.}\ \bibnamefont {Fisher}},\ }\href {\doibase 10.1103/PhysRevB.55.15832} {\bibfield  {journal} {\bibinfo  {journal} {Phys. Rev. B}\ }\textbf {\bibinfo {volume} {55}},\ \bibinfo {pages} {15832} (\bibinfo {year} {1997})}\BibitemShut {NoStop}%
\bibitem [{\citenamefont {Bid}\ \emph {et~al.}(2010)\citenamefont {Bid}, \citenamefont {Ofek}, \citenamefont {Inoue}, \citenamefont {Heiblum}, \citenamefont {Kane}, \citenamefont {Umansky},\ and\ \citenamefont {Mahalu}}]{Bid2010}%
  \BibitemOpen
  \bibfield  {author} {\bibinfo {author} {\bibfnamefont {A.}~\bibnamefont {Bid}}, \bibinfo {author} {\bibfnamefont {N.}~\bibnamefont {Ofek}}, \bibinfo {author} {\bibfnamefont {H.}~\bibnamefont {Inoue}}, \bibinfo {author} {\bibfnamefont {M.}~\bibnamefont {Heiblum}}, \bibinfo {author} {\bibfnamefont {C.~L.}\ \bibnamefont {Kane}}, \bibinfo {author} {\bibfnamefont {V.}~\bibnamefont {Umansky}}, \ and\ \bibinfo {author} {\bibfnamefont {D.}~\bibnamefont {Mahalu}},\ }\href {\doibase 10.1038/nature09277} {\bibfield  {journal} {\bibinfo  {journal} {Nature}\ }\textbf {\bibinfo {volume} {466}},\ \bibinfo {pages} {585} (\bibinfo {year} {2010})}\BibitemShut {NoStop}%
\bibitem [{\citenamefont {Takei}\ and\ \citenamefont {Rosenow}(2011)}]{Takei2011}%
  \BibitemOpen
  \bibfield  {author} {\bibinfo {author} {\bibfnamefont {S.}~\bibnamefont {Takei}}\ and\ \bibinfo {author} {\bibfnamefont {B.}~\bibnamefont {Rosenow}},\ }\href {\doibase 10.1103/PhysRevB.84.235316} {\bibfield  {journal} {\bibinfo  {journal} {Phys. Rev. B}\ }\textbf {\bibinfo {volume} {84}},\ \bibinfo {pages} {235316} (\bibinfo {year} {2011})}\BibitemShut {NoStop}%
\bibitem [{\citenamefont {Takei}\ \emph {et~al.}(2015)\citenamefont {Takei}, \citenamefont {Rosenow},\ and\ \citenamefont {Stern}}]{Takei2015}%
  \BibitemOpen
  \bibfield  {author} {\bibinfo {author} {\bibfnamefont {S.}~\bibnamefont {Takei}}, \bibinfo {author} {\bibfnamefont {B.}~\bibnamefont {Rosenow}}, \ and\ \bibinfo {author} {\bibfnamefont {A.}~\bibnamefont {Stern}},\ }\href {\doibase 10.1103/PhysRevB.91.241104} {\bibfield  {journal} {\bibinfo  {journal} {Phys. Rev. B}\ }\textbf {\bibinfo {volume} {91}},\ \bibinfo {pages} {241104} (\bibinfo {year} {2015})}\BibitemShut {NoStop}%
\bibitem [{\citenamefont {Protopopov}\ \emph {et~al.}(2017)\citenamefont {Protopopov}, \citenamefont {Gefen},\ and\ \citenamefont {Mirlin}}]{Protopopov2017}%
  \BibitemOpen
  \bibfield  {author} {\bibinfo {author} {\bibfnamefont {I.}~\bibnamefont {Protopopov}}, \bibinfo {author} {\bibfnamefont {Y.}~\bibnamefont {Gefen}}, \ and\ \bibinfo {author} {\bibfnamefont {A.}~\bibnamefont {Mirlin}},\ }\href {\doibase https://doi.org/10.1016/j.aop.2017.07.015} {\bibfield  {journal} {\bibinfo  {journal} {Annals of Physics}\ }\textbf {\bibinfo {volume} {385}},\ \bibinfo {pages} {287} (\bibinfo {year} {2017})}\BibitemShut {NoStop}%
\bibitem [{\citenamefont {Nosiglia}\ \emph {et~al.}(2018)\citenamefont {Nosiglia}, \citenamefont {Park}, \citenamefont {Rosenow},\ and\ \citenamefont {Gefen}}]{Nosiglia2018}%
  \BibitemOpen
  \bibfield  {author} {\bibinfo {author} {\bibfnamefont {C.}~\bibnamefont {Nosiglia}}, \bibinfo {author} {\bibfnamefont {J.}~\bibnamefont {Park}}, \bibinfo {author} {\bibfnamefont {B.}~\bibnamefont {Rosenow}}, \ and\ \bibinfo {author} {\bibfnamefont {Y.}~\bibnamefont {Gefen}},\ }\href {\doibase 10.1103/PhysRevB.98.115408} {\bibfield  {journal} {\bibinfo  {journal} {Phys. Rev. B}\ }\textbf {\bibinfo {volume} {98}},\ \bibinfo {pages} {115408} (\bibinfo {year} {2018})}\BibitemShut {NoStop}%
\bibitem [{\citenamefont {Park}\ \emph {et~al.}(2019)\citenamefont {Park}, \citenamefont {Mirlin}, \citenamefont {Rosenow},\ and\ \citenamefont {Gefen}}]{Park2019}%
  \BibitemOpen
  \bibfield  {author} {\bibinfo {author} {\bibfnamefont {J.}~\bibnamefont {Park}}, \bibinfo {author} {\bibfnamefont {A.~D.}\ \bibnamefont {Mirlin}}, \bibinfo {author} {\bibfnamefont {B.}~\bibnamefont {Rosenow}}, \ and\ \bibinfo {author} {\bibfnamefont {Y.}~\bibnamefont {Gefen}},\ }\href {\doibase 10.1103/PhysRevB.99.161302} {\bibfield  {journal} {\bibinfo  {journal} {Phys. Rev. B}\ }\textbf {\bibinfo {volume} {99}},\ \bibinfo {pages} {161302} (\bibinfo {year} {2019})}\BibitemShut {NoStop}%
\bibitem [{\citenamefont {Sp\aa{}nsl\"att}\ \emph {et~al.}(2020)\citenamefont {Sp\aa{}nsl\"att}, \citenamefont {Park}, \citenamefont {Gefen},\ and\ \citenamefont {Mirlin}}]{Spaanslatt2020}%
  \BibitemOpen
  \bibfield  {author} {\bibinfo {author} {\bibfnamefont {C.}~\bibnamefont {Sp\aa{}nsl\"att}}, \bibinfo {author} {\bibfnamefont {J.}~\bibnamefont {Park}}, \bibinfo {author} {\bibfnamefont {Y.}~\bibnamefont {Gefen}}, \ and\ \bibinfo {author} {\bibfnamefont {A.~D.}\ \bibnamefont {Mirlin}},\ }\href {\doibase 10.1103/PhysRevB.101.075308} {\bibfield  {journal} {\bibinfo  {journal} {Phys. Rev. B}\ }\textbf {\bibinfo {volume} {101}},\ \bibinfo {pages} {075308} (\bibinfo {year} {2020})}\BibitemShut {NoStop}%
\bibitem [{\citenamefont {Fujisawa}\ and\ \citenamefont {Lin}(2021)}]{Fujisawa2021}%
  \BibitemOpen
  \bibfield  {author} {\bibinfo {author} {\bibfnamefont {T.}~\bibnamefont {Fujisawa}}\ and\ \bibinfo {author} {\bibfnamefont {C.}~\bibnamefont {Lin}},\ }\href {\doibase 10.1103/PhysRevB.103.165302} {\bibfield  {journal} {\bibinfo  {journal} {Phys. Rev. B}\ }\textbf {\bibinfo {volume} {103}},\ \bibinfo {pages} {165302} (\bibinfo {year} {2021})}\BibitemShut {NoStop}%
\bibitem [{\citenamefont {Ponomarenko}\ and\ \citenamefont {Lyanda-Geller}(2023)}]{Ponomarenko2023}%
  \BibitemOpen
  \bibfield  {author} {\bibinfo {author} {\bibfnamefont {V.}~\bibnamefont {Ponomarenko}}\ and\ \bibinfo {author} {\bibfnamefont {Y.}~\bibnamefont {Lyanda-Geller}},\ }\href@noop {} {\enquote {\bibinfo {title} {Unusual quasiparticles and tunneling conductance in quantum point contacts in $\nu=2/3$ fractional quantum hall systems},}\ } (\bibinfo {year} {2023}),\ \Eprint {http://arxiv.org/abs/cond-mat.mes-hall/2311.05142} {arXiv:cond-mat.mes-hall/2311.05142} \BibitemShut {NoStop}%
\bibitem [{\citenamefont {Manna}\ \emph {et~al.}(2023)\citenamefont {Manna}, \citenamefont {Das},\ and\ \citenamefont {Goldstein}}]{Manna2023}%
  \BibitemOpen
  \bibfield  {author} {\bibinfo {author} {\bibfnamefont {S.}~\bibnamefont {Manna}}, \bibinfo {author} {\bibfnamefont {A.}~\bibnamefont {Das}}, \ and\ \bibinfo {author} {\bibfnamefont {M.}~\bibnamefont {Goldstein}},\ }\href@noop {} {\enquote {\bibinfo {title} {Shot noise classification of different conductance plateaus in a quantum point contact at the $\nu=2/3$ edge},}\ } (\bibinfo {year} {2023}),\ \Eprint {http://arxiv.org/abs/cond-mat.mes-hall/2307.05175} {arXiv:cond-mat.mes-hall/2307.05175 [cond-mat.mes-hall]} \BibitemShut {NoStop}%
\bibitem [{\citenamefont {Meir}(1994)}]{Meir1994}%
  \BibitemOpen
  \bibfield  {author} {\bibinfo {author} {\bibfnamefont {Y.}~\bibnamefont {Meir}},\ }\href {\doibase 10.1103/PhysRevLett.72.2624} {\bibfield  {journal} {\bibinfo  {journal} {Phys. Rev. Lett.}\ }\textbf {\bibinfo {volume} {72}},\ \bibinfo {pages} {2624} (\bibinfo {year} {1994})}\BibitemShut {NoStop}%
\bibitem [{\citenamefont {Wang}\ \emph {et~al.}(2013)\citenamefont {Wang}, \citenamefont {Meir},\ and\ \citenamefont {Gefen}}]{Wang2013}%
  \BibitemOpen
  \bibfield  {author} {\bibinfo {author} {\bibfnamefont {J.}~\bibnamefont {Wang}}, \bibinfo {author} {\bibfnamefont {Y.}~\bibnamefont {Meir}}, \ and\ \bibinfo {author} {\bibfnamefont {Y.}~\bibnamefont {Gefen}},\ }\href {\doibase 10.1103/PhysRevLett.111.246803} {\bibfield  {journal} {\bibinfo  {journal} {Phys. Rev. Lett.}\ }\textbf {\bibinfo {volume} {111}},\ \bibinfo {pages} {246803} (\bibinfo {year} {2013})}\BibitemShut {NoStop}%
\bibitem [{\citenamefont {Gross}\ \emph {et~al.}(2012)\citenamefont {Gross}, \citenamefont {Dolev}, \citenamefont {Heiblum}, \citenamefont {Umansky},\ and\ \citenamefont {Mahalu}}]{Gross2012}%
  \BibitemOpen
  \bibfield  {author} {\bibinfo {author} {\bibfnamefont {Y.}~\bibnamefont {Gross}}, \bibinfo {author} {\bibfnamefont {M.}~\bibnamefont {Dolev}}, \bibinfo {author} {\bibfnamefont {M.}~\bibnamefont {Heiblum}}, \bibinfo {author} {\bibfnamefont {V.}~\bibnamefont {Umansky}}, \ and\ \bibinfo {author} {\bibfnamefont {D.}~\bibnamefont {Mahalu}},\ }\href {\doibase 10.1103/PhysRevLett.108.226801} {\bibfield  {journal} {\bibinfo  {journal} {Phys. Rev. Lett.}\ }\textbf {\bibinfo {volume} {108}},\ \bibinfo {pages} {226801} (\bibinfo {year} {2012})}\BibitemShut {NoStop}%
\bibitem [{\citenamefont {Altimiras}\ \emph {et~al.}(2012)\citenamefont {Altimiras}, \citenamefont {le~Sueur}, \citenamefont {Gennser}, \citenamefont {Anthore}, \citenamefont {Cavanna}, \citenamefont {Mailly},\ and\ \citenamefont {Pierre}}]{Altimiras2012}%
  \BibitemOpen
  \bibfield  {author} {\bibinfo {author} {\bibfnamefont {C.}~\bibnamefont {Altimiras}}, \bibinfo {author} {\bibfnamefont {H.}~\bibnamefont {le~Sueur}}, \bibinfo {author} {\bibfnamefont {U.}~\bibnamefont {Gennser}}, \bibinfo {author} {\bibfnamefont {A.}~\bibnamefont {Anthore}}, \bibinfo {author} {\bibfnamefont {A.}~\bibnamefont {Cavanna}}, \bibinfo {author} {\bibfnamefont {D.}~\bibnamefont {Mailly}}, \ and\ \bibinfo {author} {\bibfnamefont {F.}~\bibnamefont {Pierre}},\ }\href {\doibase 10.1103/PhysRevLett.109.026803} {\bibfield  {journal} {\bibinfo  {journal} {Phys. Rev. Lett.}\ }\textbf {\bibinfo {volume} {109}},\ \bibinfo {pages} {026803} (\bibinfo {year} {2012})}\BibitemShut {NoStop}%
\bibitem [{\citenamefont {Inoue}\ \emph {et~al.}(2014)\citenamefont {Inoue}, \citenamefont {Grivnin}, \citenamefont {Ronen}, \citenamefont {Heiblum}, \citenamefont {Umansky},\ and\ \citenamefont {Mahalu}}]{Inoue2014}%
  \BibitemOpen
  \bibfield  {author} {\bibinfo {author} {\bibfnamefont {H.}~\bibnamefont {Inoue}}, \bibinfo {author} {\bibfnamefont {A.}~\bibnamefont {Grivnin}}, \bibinfo {author} {\bibfnamefont {Y.}~\bibnamefont {Ronen}}, \bibinfo {author} {\bibfnamefont {M.}~\bibnamefont {Heiblum}}, \bibinfo {author} {\bibfnamefont {V.}~\bibnamefont {Umansky}}, \ and\ \bibinfo {author} {\bibfnamefont {D.}~\bibnamefont {Mahalu}},\ }\href {\doibase 10.1038/ncomms5067} {\bibfield  {journal} {\bibinfo  {journal} {Nature Communications}\ }\textbf {\bibinfo {volume} {5}},\ \bibinfo {pages} {4067} (\bibinfo {year} {2014})}\BibitemShut {NoStop}%
\bibitem [{\citenamefont {Banerjee}\ \emph {et~al.}(2017)\citenamefont {Banerjee}, \citenamefont {Heiblum}, \citenamefont {Rosenblatt}, \citenamefont {Oreg}, \citenamefont {Feldman}, \citenamefont {Stern},\ and\ \citenamefont {Umansky}}]{Banerjee2017}%
  \BibitemOpen
  \bibfield  {author} {\bibinfo {author} {\bibfnamefont {M.}~\bibnamefont {Banerjee}}, \bibinfo {author} {\bibfnamefont {M.}~\bibnamefont {Heiblum}}, \bibinfo {author} {\bibfnamefont {A.}~\bibnamefont {Rosenblatt}}, \bibinfo {author} {\bibfnamefont {Y.}~\bibnamefont {Oreg}}, \bibinfo {author} {\bibfnamefont {D.~E.}\ \bibnamefont {Feldman}}, \bibinfo {author} {\bibfnamefont {A.}~\bibnamefont {Stern}}, \ and\ \bibinfo {author} {\bibfnamefont {V.}~\bibnamefont {Umansky}},\ }\href {\doibase 10.1038/nature22052} {\bibfield  {journal} {\bibinfo  {journal} {Nature}\ }\textbf {\bibinfo {volume} {545}},\ \bibinfo {pages} {75} (\bibinfo {year} {2017})}\BibitemShut {NoStop}%
\bibitem [{\citenamefont {Rosenblatt}\ \emph {et~al.}(2020)\citenamefont {Rosenblatt}, \citenamefont {Konyzheva}, \citenamefont {Lafont}, \citenamefont {Schiller}, \citenamefont {Park}, \citenamefont {Snizhko}, \citenamefont {Heiblum}, \citenamefont {Oreg},\ and\ \citenamefont {Umansky}}]{Rosenblatt2020}%
  \BibitemOpen
  \bibfield  {author} {\bibinfo {author} {\bibfnamefont {A.}~\bibnamefont {Rosenblatt}}, \bibinfo {author} {\bibfnamefont {S.}~\bibnamefont {Konyzheva}}, \bibinfo {author} {\bibfnamefont {F.}~\bibnamefont {Lafont}}, \bibinfo {author} {\bibfnamefont {N.}~\bibnamefont {Schiller}}, \bibinfo {author} {\bibfnamefont {J.}~\bibnamefont {Park}}, \bibinfo {author} {\bibfnamefont {K.}~\bibnamefont {Snizhko}}, \bibinfo {author} {\bibfnamefont {M.}~\bibnamefont {Heiblum}}, \bibinfo {author} {\bibfnamefont {Y.}~\bibnamefont {Oreg}}, \ and\ \bibinfo {author} {\bibfnamefont {V.}~\bibnamefont {Umansky}},\ }\href {\doibase 10.1103/PhysRevLett.125.256803} {\bibfield  {journal} {\bibinfo  {journal} {Phys. Rev. Lett.}\ }\textbf {\bibinfo {volume} {125}},\ \bibinfo {pages} {256803} (\bibinfo {year} {2020})}\BibitemShut {NoStop}%
\bibitem [{\citenamefont {Srivastav}\ \emph {et~al.}(2021)\citenamefont {Srivastav}, \citenamefont {Kumar}, \citenamefont {Sp\aa{}nsl\"att}, \citenamefont {Watanabe}, \citenamefont {Taniguchi}, \citenamefont {Mirlin}, \citenamefont {Gefen},\ and\ \citenamefont {Das}}]{Srivastav2021}%
  \BibitemOpen
  \bibfield  {author} {\bibinfo {author} {\bibfnamefont {S.~K.}\ \bibnamefont {Srivastav}}, \bibinfo {author} {\bibfnamefont {R.}~\bibnamefont {Kumar}}, \bibinfo {author} {\bibfnamefont {C.}~\bibnamefont {Sp\aa{}nsl\"att}}, \bibinfo {author} {\bibfnamefont {K.}~\bibnamefont {Watanabe}}, \bibinfo {author} {\bibfnamefont {T.}~\bibnamefont {Taniguchi}}, \bibinfo {author} {\bibfnamefont {A.~D.}\ \bibnamefont {Mirlin}}, \bibinfo {author} {\bibfnamefont {Y.}~\bibnamefont {Gefen}}, \ and\ \bibinfo {author} {\bibfnamefont {A.}~\bibnamefont {Das}},\ }\href {\doibase 10.1103/PhysRevLett.126.216803} {\bibfield  {journal} {\bibinfo  {journal} {Phys. Rev. Lett.}\ }\textbf {\bibinfo {volume} {126}},\ \bibinfo {pages} {216803} (\bibinfo {year} {2021})}\BibitemShut {NoStop}%
\bibitem [{\citenamefont {Le~Breton}\ \emph {et~al.}(2022)\citenamefont {Le~Breton}, \citenamefont {Delagrange}, \citenamefont {Hong}, \citenamefont {Garg}, \citenamefont {Watanabe}, \citenamefont {Taniguchi}, \citenamefont {Ribeiro-Palau}, \citenamefont {Roulleau}, \citenamefont {Roche},\ and\ \citenamefont {Parmentier}}]{LeBreton2022}%
  \BibitemOpen
  \bibfield  {author} {\bibinfo {author} {\bibfnamefont {G.}~\bibnamefont {Le~Breton}}, \bibinfo {author} {\bibfnamefont {R.}~\bibnamefont {Delagrange}}, \bibinfo {author} {\bibfnamefont {Y.}~\bibnamefont {Hong}}, \bibinfo {author} {\bibfnamefont {M.}~\bibnamefont {Garg}}, \bibinfo {author} {\bibfnamefont {K.}~\bibnamefont {Watanabe}}, \bibinfo {author} {\bibfnamefont {T.}~\bibnamefont {Taniguchi}}, \bibinfo {author} {\bibfnamefont {R.}~\bibnamefont {Ribeiro-Palau}}, \bibinfo {author} {\bibfnamefont {P.}~\bibnamefont {Roulleau}}, \bibinfo {author} {\bibfnamefont {P.}~\bibnamefont {Roche}}, \ and\ \bibinfo {author} {\bibfnamefont {F.~D.}\ \bibnamefont {Parmentier}},\ }\href {\doibase 10.1103/PhysRevLett.129.116803} {\bibfield  {journal} {\bibinfo  {journal} {Phys. Rev. Lett.}\ }\textbf {\bibinfo {volume} {129}},\ \bibinfo {pages} {116803} (\bibinfo {year} {2022})}\BibitemShut {NoStop}%
\bibitem [{\citenamefont {Nakamura}\ \emph {et~al.}(2023)\citenamefont {Nakamura}, \citenamefont {Liang}, \citenamefont {Gardner},\ and\ \citenamefont {Manfra}}]{Nakamura2023}%
  \BibitemOpen
  \bibfield  {author} {\bibinfo {author} {\bibfnamefont {J.}~\bibnamefont {Nakamura}}, \bibinfo {author} {\bibfnamefont {S.}~\bibnamefont {Liang}}, \bibinfo {author} {\bibfnamefont {G.~C.}\ \bibnamefont {Gardner}}, \ and\ \bibinfo {author} {\bibfnamefont {M.~J.}\ \bibnamefont {Manfra}},\ }\href {\doibase 10.1103/PhysRevLett.130.076205} {\bibfield  {journal} {\bibinfo  {journal} {Phys. Rev. Lett.}\ }\textbf {\bibinfo {volume} {130}},\ \bibinfo {pages} {076205} (\bibinfo {year} {2023})}\BibitemShut {NoStop}%
\bibitem [{\citenamefont {Hashisaka}\ \emph {et~al.}(2023)\citenamefont {Hashisaka}, \citenamefont {Ito}, \citenamefont {Akiho}, \citenamefont {Sasaki}, \citenamefont {Kumada}, \citenamefont {Shibata},\ and\ \citenamefont {Muraki}}]{Hashisaka2023}%
  \BibitemOpen
  \bibfield  {author} {\bibinfo {author} {\bibfnamefont {M.}~\bibnamefont {Hashisaka}}, \bibinfo {author} {\bibfnamefont {T.}~\bibnamefont {Ito}}, \bibinfo {author} {\bibfnamefont {T.}~\bibnamefont {Akiho}}, \bibinfo {author} {\bibfnamefont {S.}~\bibnamefont {Sasaki}}, \bibinfo {author} {\bibfnamefont {N.}~\bibnamefont {Kumada}}, \bibinfo {author} {\bibfnamefont {N.}~\bibnamefont {Shibata}}, \ and\ \bibinfo {author} {\bibfnamefont {K.}~\bibnamefont {Muraki}},\ }\href {\doibase 10.1103/PhysRevX.13.031024} {\bibfield  {journal} {\bibinfo  {journal} {Phys. Rev. X}\ }\textbf {\bibinfo {volume} {13}},\ \bibinfo {pages} {031024} (\bibinfo {year} {2023})}\BibitemShut {NoStop}%
\bibitem [{\citenamefont {Cohen}\ \emph {et~al.}(2019)\citenamefont {Cohen}, \citenamefont {Ronen}, \citenamefont {Yang}, \citenamefont {Banitt}, \citenamefont {Park}, \citenamefont {Heiblum}, \citenamefont {Mirlin}, \citenamefont {Gefen},\ and\ \citenamefont {Umansky}}]{Cohen2019}%
  \BibitemOpen
  \bibfield  {author} {\bibinfo {author} {\bibfnamefont {Y.}~\bibnamefont {Cohen}}, \bibinfo {author} {\bibfnamefont {Y.}~\bibnamefont {Ronen}}, \bibinfo {author} {\bibfnamefont {W.}~\bibnamefont {Yang}}, \bibinfo {author} {\bibfnamefont {D.}~\bibnamefont {Banitt}}, \bibinfo {author} {\bibfnamefont {J.}~\bibnamefont {Park}}, \bibinfo {author} {\bibfnamefont {M.}~\bibnamefont {Heiblum}}, \bibinfo {author} {\bibfnamefont {A.~D.}\ \bibnamefont {Mirlin}}, \bibinfo {author} {\bibfnamefont {Y.}~\bibnamefont {Gefen}}, \ and\ \bibinfo {author} {\bibfnamefont {V.}~\bibnamefont {Umansky}},\ }\href {\doibase 10.1038/s41467-019-09920-5} {\bibfield  {journal} {\bibinfo  {journal} {Nature Communications}\ }\textbf {\bibinfo {volume} {10}},\ \bibinfo {pages} {1920} (\bibinfo {year} {2019})}\BibitemShut {NoStop}%
\bibitem [{\citenamefont {Hashisaka}\ \emph {et~al.}(2021)\citenamefont {Hashisaka}, \citenamefont {Jonckheere}, \citenamefont {Akiho}, \citenamefont {Sasaki}, \citenamefont {Rech}, \citenamefont {Martin},\ and\ \citenamefont {Muraki}}]{Hashisaka2021}%
  \BibitemOpen
  \bibfield  {author} {\bibinfo {author} {\bibfnamefont {M.}~\bibnamefont {Hashisaka}}, \bibinfo {author} {\bibfnamefont {T.}~\bibnamefont {Jonckheere}}, \bibinfo {author} {\bibfnamefont {T.}~\bibnamefont {Akiho}}, \bibinfo {author} {\bibfnamefont {S.}~\bibnamefont {Sasaki}}, \bibinfo {author} {\bibfnamefont {J.}~\bibnamefont {Rech}}, \bibinfo {author} {\bibfnamefont {T.}~\bibnamefont {Martin}}, \ and\ \bibinfo {author} {\bibfnamefont {K.}~\bibnamefont {Muraki}},\ }\href {\doibase 10.1038/s41467-021-23160-6} {\bibfield  {journal} {\bibinfo  {journal} {Nature Communications}\ }\textbf {\bibinfo {volume} {12}},\ \bibinfo {pages} {2794} (\bibinfo {year} {2021})}\BibitemShut {NoStop}%
\bibitem [{\citenamefont {St\"abler}\ and\ \citenamefont {Sukhorukov}(2022)}]{Staebler2022}%
  \BibitemOpen
  \bibfield  {author} {\bibinfo {author} {\bibfnamefont {F.}~\bibnamefont {St\"abler}}\ and\ \bibinfo {author} {\bibfnamefont {E.}~\bibnamefont {Sukhorukov}},\ }\href {\doibase 10.1103/PhysRevB.105.235417} {\bibfield  {journal} {\bibinfo  {journal} {Phys. Rev. B}\ }\textbf {\bibinfo {volume} {105}},\ \bibinfo {pages} {235417} (\bibinfo {year} {2022})}\BibitemShut {NoStop}%
\bibitem [{\citenamefont {St\"abler}\ and\ \citenamefont {Sukhorukov}(2023)}]{Staebler2023}%
  \BibitemOpen
  \bibfield  {author} {\bibinfo {author} {\bibfnamefont {F.}~\bibnamefont {St\"abler}}\ and\ \bibinfo {author} {\bibfnamefont {E.}~\bibnamefont {Sukhorukov}},\ }\href {\doibase 10.1103/PhysRevB.107.045403} {\bibfield  {journal} {\bibinfo  {journal} {Phys. Rev. B}\ }\textbf {\bibinfo {volume} {107}},\ \bibinfo {pages} {045403} (\bibinfo {year} {2023})}\BibitemShut {NoStop}%
\bibitem [{\citenamefont {{de Jong}}\ and\ \citenamefont {Beenakker}(1996)}]{deJong1996}%
  \BibitemOpen
  \bibfield  {author} {\bibinfo {author} {\bibfnamefont {M.}~\bibnamefont {{de Jong}}}\ and\ \bibinfo {author} {\bibfnamefont {C.}~\bibnamefont {Beenakker}},\ }\href {\doibase https://doi.org/10.1016/0378-4371(96)00068-4} {\bibfield  {journal} {\bibinfo  {journal} {Physica A}\ }\textbf {\bibinfo {volume} {230}},\ \bibinfo {pages} {219} (\bibinfo {year} {1996})}\BibitemShut {NoStop}%
\bibitem [{\citenamefont {Acciai}\ \emph {et~al.}(2022)\citenamefont {Acciai}, \citenamefont {Roulleau}, \citenamefont {Taktak}, \citenamefont {Glattli},\ and\ \citenamefont {Splettstoesser}}]{Acciai2022}%
  \BibitemOpen
  \bibfield  {author} {\bibinfo {author} {\bibfnamefont {M.}~\bibnamefont {Acciai}}, \bibinfo {author} {\bibfnamefont {P.}~\bibnamefont {Roulleau}}, \bibinfo {author} {\bibfnamefont {I.}~\bibnamefont {Taktak}}, \bibinfo {author} {\bibfnamefont {D.~C.}\ \bibnamefont {Glattli}}, \ and\ \bibinfo {author} {\bibfnamefont {J.}~\bibnamefont {Splettstoesser}},\ }\href {\doibase 10.1103/PhysRevB.105.125415} {\bibfield  {journal} {\bibinfo  {journal} {Phys. Rev. B}\ }\textbf {\bibinfo {volume} {105}},\ \bibinfo {pages} {125415} (\bibinfo {year} {2022})}\BibitemShut {NoStop}%
\bibitem [{\citenamefont {Sp\aa{}nsl\"att}\ \emph {et~al.}(2019)\citenamefont {Sp\aa{}nsl\"att}, \citenamefont {Park}, \citenamefont {Gefen},\ and\ \citenamefont {Mirlin}}]{Park2019-b}%
  \BibitemOpen
  \bibfield  {author} {\bibinfo {author} {\bibfnamefont {C.}~\bibnamefont {Sp\aa{}nsl\"att}}, \bibinfo {author} {\bibfnamefont {J.}~\bibnamefont {Park}}, \bibinfo {author} {\bibfnamefont {Y.}~\bibnamefont {Gefen}}, \ and\ \bibinfo {author} {\bibfnamefont {A.~D.}\ \bibnamefont {Mirlin}},\ }\href {\doibase 10.1103/PhysRevLett.123.137701} {\bibfield  {journal} {\bibinfo  {journal} {Phys. Rev. Lett.}\ }\textbf {\bibinfo {volume} {123}},\ \bibinfo {pages} {137701} (\bibinfo {year} {2019})}\BibitemShut {NoStop}%
\bibitem [{\citenamefont {Kumar}\ \emph {et~al.}(2022)\citenamefont {Kumar}, \citenamefont {Srivastav}, \citenamefont {Sp{\aa}nsl{\"a}tt}, \citenamefont {Watanabe}, \citenamefont {Taniguchi}, \citenamefont {Gefen}, \citenamefont {Mirlin},\ and\ \citenamefont {Das}}]{Kumar2022}%
  \BibitemOpen
  \bibfield  {author} {\bibinfo {author} {\bibfnamefont {R.}~\bibnamefont {Kumar}}, \bibinfo {author} {\bibfnamefont {S.~K.}\ \bibnamefont {Srivastav}}, \bibinfo {author} {\bibfnamefont {C.}~\bibnamefont {Sp{\aa}nsl{\"a}tt}}, \bibinfo {author} {\bibfnamefont {K.}~\bibnamefont {Watanabe}}, \bibinfo {author} {\bibfnamefont {T.}~\bibnamefont {Taniguchi}}, \bibinfo {author} {\bibfnamefont {Y.}~\bibnamefont {Gefen}}, \bibinfo {author} {\bibfnamefont {A.~D.}\ \bibnamefont {Mirlin}}, \ and\ \bibinfo {author} {\bibfnamefont {A.}~\bibnamefont {Das}},\ }\href {\doibase 10.1038/s41467-021-27805-4} {\bibfield  {journal} {\bibinfo  {journal} {Nature Communications}\ }\textbf {\bibinfo {volume} {13}},\ \bibinfo {pages} {213} (\bibinfo {year} {2022})}\BibitemShut {NoStop}%
\bibitem [{\citenamefont {Melcer}\ \emph {et~al.}(2022)\citenamefont {Melcer}, \citenamefont {Dutta}, \citenamefont {Spånslätt}, \citenamefont {Park}, \citenamefont {Mirlin},\ and\ \citenamefont {Umansky}}]{Melcer2022}%
  \BibitemOpen
  \bibfield  {author} {\bibinfo {author} {\bibfnamefont {R.~A.}\ \bibnamefont {Melcer}}, \bibinfo {author} {\bibfnamefont {B.}~\bibnamefont {Dutta}}, \bibinfo {author} {\bibfnamefont {C.}~\bibnamefont {Spånslätt}}, \bibinfo {author} {\bibfnamefont {J.}~\bibnamefont {Park}}, \bibinfo {author} {\bibfnamefont {A.~D.}\ \bibnamefont {Mirlin}}, \ and\ \bibinfo {author} {\bibfnamefont {V.}~\bibnamefont {Umansky}},\ }\href {\doibase 10.1038/s41467-022-28009-0} {\bibfield  {journal} {\bibinfo  {journal} {Nature Communications}\ }\textbf {\bibinfo {volume} {13}},\ \bibinfo {pages} {376} (\bibinfo {year} {2022})}\BibitemShut {NoStop}%
\bibitem [{\citenamefont {Manna}\ \emph {et~al.}(2024)\citenamefont {Manna}, \citenamefont {Das}, \citenamefont {Gefen},\ and\ \citenamefont {Goldstein}}]{Manna2024}%
  \BibitemOpen
  \bibfield  {author} {\bibinfo {author} {\bibfnamefont {S.}~\bibnamefont {Manna}}, \bibinfo {author} {\bibfnamefont {A.}~\bibnamefont {Das}}, \bibinfo {author} {\bibfnamefont {Y.}~\bibnamefont {Gefen}}, \ and\ \bibinfo {author} {\bibfnamefont {M.}~\bibnamefont {Goldstein}},\ }\href@noop {} {\enquote {\bibinfo {title} {Diagnostics of anomalous conductance plateaus in abelian quantum hall regime},}\ } (\bibinfo {year} {2024}),\ \Eprint {http://arxiv.org/abs/preprint} {arXiv:preprint [cond-mat.mes-hall]} \BibitemShut {NoStop}%
\bibitem [{\citenamefont {Lumbroso}\ \emph {et~al.}(2018)\citenamefont {Lumbroso}, \citenamefont {Simine}, \citenamefont {Nitzan}, \citenamefont {Segal},\ and\ \citenamefont {Tal}}]{Lumbroso2018}%
  \BibitemOpen
  \bibfield  {author} {\bibinfo {author} {\bibfnamefont {O.~S.}\ \bibnamefont {Lumbroso}}, \bibinfo {author} {\bibfnamefont {L.}~\bibnamefont {Simine}}, \bibinfo {author} {\bibfnamefont {A.}~\bibnamefont {Nitzan}}, \bibinfo {author} {\bibfnamefont {D.}~\bibnamefont {Segal}}, \ and\ \bibinfo {author} {\bibfnamefont {O.}~\bibnamefont {Tal}},\ }\href {\doibase 10.1038/s41586-018-0592-2} {\bibfield  {journal} {\bibinfo  {journal} {Nature}\ }\textbf {\bibinfo {volume} {562}},\ \bibinfo {pages} {240} (\bibinfo {year} {2018})}\BibitemShut {NoStop}%
\bibitem [{\citenamefont {Larocque}\ \emph {et~al.}(2020)\citenamefont {Larocque}, \citenamefont {Pinsolle}, \citenamefont {Lupien},\ and\ \citenamefont {Reulet}}]{Reulet2020}%
  \BibitemOpen
  \bibfield  {author} {\bibinfo {author} {\bibfnamefont {S.}~\bibnamefont {Larocque}}, \bibinfo {author} {\bibfnamefont {E.}~\bibnamefont {Pinsolle}}, \bibinfo {author} {\bibfnamefont {C.}~\bibnamefont {Lupien}}, \ and\ \bibinfo {author} {\bibfnamefont {B.}~\bibnamefont {Reulet}},\ }\href {\doibase 10.1103/PhysRevLett.125.106801} {\bibfield  {journal} {\bibinfo  {journal} {Phys. Rev. Lett.}\ }\textbf {\bibinfo {volume} {125}},\ \bibinfo {pages} {106801} (\bibinfo {year} {2020})}\BibitemShut {NoStop}%
\bibitem [{\citenamefont {Rech}\ \emph {et~al.}(2020)\citenamefont {Rech}, \citenamefont {Jonckheere}, \citenamefont {Gr\'emaud},\ and\ \citenamefont {Martin}}]{Rech2020}%
  \BibitemOpen
  \bibfield  {author} {\bibinfo {author} {\bibfnamefont {J.}~\bibnamefont {Rech}}, \bibinfo {author} {\bibfnamefont {T.}~\bibnamefont {Jonckheere}}, \bibinfo {author} {\bibfnamefont {B.}~\bibnamefont {Gr\'emaud}}, \ and\ \bibinfo {author} {\bibfnamefont {T.}~\bibnamefont {Martin}},\ }\href {\doibase 10.1103/PhysRevLett.125.086801} {\bibfield  {journal} {\bibinfo  {journal} {Phys. Rev. Lett.}\ }\textbf {\bibinfo {volume} {125}},\ \bibinfo {pages} {086801} (\bibinfo {year} {2020})}\BibitemShut {NoStop}%
\bibitem [{\citenamefont {Rebora}\ \emph {et~al.}(2022)\citenamefont {Rebora}, \citenamefont {Rech}, \citenamefont {Ferraro}, \citenamefont {Jonckheere}, \citenamefont {Martin},\ and\ \citenamefont {Sassetti}}]{Sassetti2022}%
  \BibitemOpen
  \bibfield  {author} {\bibinfo {author} {\bibfnamefont {G.}~\bibnamefont {Rebora}}, \bibinfo {author} {\bibfnamefont {J.}~\bibnamefont {Rech}}, \bibinfo {author} {\bibfnamefont {D.}~\bibnamefont {Ferraro}}, \bibinfo {author} {\bibfnamefont {T.}~\bibnamefont {Jonckheere}}, \bibinfo {author} {\bibfnamefont {T.}~\bibnamefont {Martin}}, \ and\ \bibinfo {author} {\bibfnamefont {M.}~\bibnamefont {Sassetti}},\ }\href {\doibase 10.1103/PhysRevResearch.4.043191} {\bibfield  {journal} {\bibinfo  {journal} {Phys. Rev. Res.}\ }\textbf {\bibinfo {volume} {4}},\ \bibinfo {pages} {043191} (\bibinfo {year} {2022})}\BibitemShut {NoStop}%
\bibitem [{\citenamefont {Acciai}\ \emph {et~al.}(2024)\citenamefont {Acciai}, \citenamefont {Sp\aa{}nsl\"att},\ and\ \citenamefont {Zhang}}]{Acciai2024}%
  \BibitemOpen
  \bibfield  {author} {\bibinfo {author} {\bibfnamefont {M.}~\bibnamefont {Acciai}}, \bibinfo {author} {\bibfnamefont {C.}~\bibnamefont {Sp\aa{}nsl\"att}}, \ and\ \bibinfo {author} {\bibfnamefont {G.}~\bibnamefont {Zhang}},\ }\href@noop {} {\enquote {\bibinfo {title} {Role of scaling dimensions in generalized noises in fractional quantum hall tunneling due to a temperature bias},}\ } (\bibinfo {year} {2024}),\ \Eprint {http://arxiv.org/abs/cond-mat.mes-hall/2408.04525} {arXiv:cond-mat.mes-hall/2408.04525} \BibitemShut {NoStop}%
\bibitem [{\citenamefont {Lin}\ \emph {et~al.}(2021)\citenamefont {Lin}, \citenamefont {Hashisaka}, \citenamefont {Akiho}, \citenamefont {Muraki},\ and\ \citenamefont {Fujisawa}}]{Lin2021}%
  \BibitemOpen
  \bibfield  {author} {\bibinfo {author} {\bibfnamefont {C.}~\bibnamefont {Lin}}, \bibinfo {author} {\bibfnamefont {M.}~\bibnamefont {Hashisaka}}, \bibinfo {author} {\bibfnamefont {T.}~\bibnamefont {Akiho}}, \bibinfo {author} {\bibfnamefont {K.}~\bibnamefont {Muraki}}, \ and\ \bibinfo {author} {\bibfnamefont {T.}~\bibnamefont {Fujisawa}},\ }\href {\doibase 10.1103/PhysRevB.104.125304} {\bibfield  {journal} {\bibinfo  {journal} {Phys. Rev. B}\ }\textbf {\bibinfo {volume} {104}},\ \bibinfo {pages} {125304} (\bibinfo {year} {2021})}\BibitemShut {NoStop}%
\bibitem [{\citenamefont {Wen}(1991)}]{XGWen1991}%
  \BibitemOpen
  \bibfield  {author} {\bibinfo {author} {\bibfnamefont {X.-G.}\ \bibnamefont {Wen}},\ }\href {\doibase 10.1103/PhysRevB.44.5708} {\bibfield  {journal} {\bibinfo  {journal} {Phys. Rev. B}\ }\textbf {\bibinfo {volume} {44}},\ \bibinfo {pages} {5708} (\bibinfo {year} {1991})}\BibitemShut {NoStop}%
\bibitem [{\citenamefont {Kapfer}\ \emph {et~al.}(2019)\citenamefont {Kapfer}, \citenamefont {Roulleau}, \citenamefont {Santin}, \citenamefont {Farrer}, \citenamefont {Ritchie},\ and\ \citenamefont {Glattli}}]{Kapfer2019}%
  \BibitemOpen
  \bibfield  {author} {\bibinfo {author} {\bibfnamefont {M.}~\bibnamefont {Kapfer}}, \bibinfo {author} {\bibfnamefont {P.}~\bibnamefont {Roulleau}}, \bibinfo {author} {\bibfnamefont {M.}~\bibnamefont {Santin}}, \bibinfo {author} {\bibfnamefont {I.}~\bibnamefont {Farrer}}, \bibinfo {author} {\bibfnamefont {D.~A.}\ \bibnamefont {Ritchie}}, \ and\ \bibinfo {author} {\bibfnamefont {D.~C.}\ \bibnamefont {Glattli}},\ }\href {\doibase 10.1126/science.aau353} {\bibfield  {journal} {\bibinfo  {journal} {Science}\ }\textbf {\bibinfo {volume} {363}},\ \bibinfo {pages} {846} (\bibinfo {year} {2019})}\BibitemShut {NoStop}%
\bibitem [{\citenamefont {Safi}(2014)}]{safi2014}%
  \BibitemOpen
  \bibfield  {author} {\bibinfo {author} {\bibfnamefont {I.}~\bibnamefont {Safi}},\ }\href@noop {} {\enquote {\bibinfo {title} {{Time-dependent Transport in arbitrary extended driven tunnel junctions}},}\ } (\bibinfo {year} {2014}),\ \Eprint {http://arxiv.org/abs/1401.5950} {arXiv:1401.5950 [cond-mat.mes-hall]} \BibitemShut {NoStop}%
\end{thebibliography}%

\end{document}